\documentclass{aa}  
\usepackage{orcidlink}
\usepackage{graphicx}
\usepackage[outdir=./]{epstopdf}
\usepackage{txfonts}
\usepackage{multirow}
\usepackage{lscape}
\usepackage{amsmath}
\usepackage{lastpage}
\usepackage{url}
\usepackage{supertabular}

\newcommand{\rsun}{R$_{\odot}$}
\newcommand{\lsun}{L$_{\odot}$}

\newcommand{\acyg}{$\alpha$ Cygni}

\newcommand{\logt}[2]{log\,$T_{\rm eff}#1#2$}
\newcommand{\logl}[2]{log\,($L$/\lsun$)#1#2$}
\newcommand{\gammae}[2]{$\Gamma_{e}#1#2$}
\newcommand{\ared}[2]{$\alpha_{0}#1#2$}
\newcommand{\awhi}[2]{$\alpha_{w}#1#2$}

\begin{document} 

\title{Variability of Galactic blue supergiants observed with TESS}

\author{
Michalis Kourniotis\orcidlink{0000-0003-0206-8802
}\inst{1}\thanks{email: kourniotis@asu.cas.cz}
\and Lydia S. Cidale\orcidlink{0000-0003-2160-7146}\inst{2,3}
\and Michaela Kraus\orcidlink{0000-0002-4502-6330}\inst{1}
\and  Matias A. Ruiz Diaz\orcidlink{0000-0003-2384-6044}\inst{2,3}
\and Aldana Alberici Adam\orcidlink{0009-0005-3909-6761}\inst{2,3}
          }

\institute{
Astronomical Institute, Czech Academy of Sciences, Fri\v{c}ova 298, 251\,65 Ond\v{r}ejov, Czech Republic
\and
Instituto de Astrof\'{\i}sica La Plata, CCT La Plata, CONICET-UNLP, Paseo del Bosque S/N, B1900FWA, La Plata, Argentina
\and
Departamento de Espectroscop\'{\i}a, Facultad de Ciencias Astron\'omicas y Geof\'{\i}sicas, Universidad Nacional de La Plata (UNLP), Paseo del Bosque S/N, B1900FWA, La Plata, Argentina
}

\date{Received 24/09/2024; accepted 26/03/2025}
 
\abstract
   {Blue supergiants (BSGs) mediate between the main sequence and the late stages of massive stars, which makes them valuable for assessing the physics that drives the stars across the diverse evolutionary channels.} 
   {By exploring correlations between the parameters of BSGs and their variability properties, we aim to improve the constraints on the models of the evolved star structure and on the physics of the post-main-sequence evolution.}   
   {We conducted a variability study of 41 BSGs with known spectroscopic parameters in the Galaxy using high-precision photometry from the Transiting Exoplanet Survey Satellite. Stellar luminosities were calculated from the fit of multiband photometry and using the latest distance estimates from $Gaia$.  We described the time domain of the stars by means of three statistical measures and extracted prominent frequencies via an iterative pre-whitening process. Alongside, we investigated the debated stochastic low-frequency (SLF) variability, which manifests itself in all amplitude spectra. }
  {We report a positive correlation between the amplitude of photometric variability and the stellar luminosity.  For \logl{\lesssim}{5}, stars display frequencies that match the rotational one, suggesting that variability is presumably driven by surface spots and/or features embedded in the wind. For \logl{\gtrsim}{5}, variables of the \acyg~class manifest themselves with their diverse and/or time-variant photometric properties and their systematically lower frequencies. Moreover, we report a positive correlation between the SLF variability amplitude and the effective temperature, indicating an influential role that the stellar age plays on the emergence of the background signal beyond the main sequence. Positive, though weak, correlation is also observed between the intrinsic brightness and the SLF variability amplitude, similar to the findings in the Large Magellanic Cloud, which suggests an in common excitation mechanism that depends only mildly on metallicity. Exceptionally, the \acyg~variables display a suppressed SLF variability that prompts to the interior changes that the evolving stars undergo.}
  {}


\maketitle

\section{Introduction}
\label{sec:intro}

Being among the brightest stars in galaxies and young star associations, the blue supergiants (BSGs) are messengers of information on key parameters of their host environment such as age, kinematics, metallicity, and distance  \citep{2006ApJ...648.1007B,2008ApJ...684..118U,2012ApJ...747...15K}. Their accessibility enables studies to be undertaken with high resolution, and on a systematic basis, providing the means to describe mechanisms that are pivotal to the evolution of massive stars.

The high number of reported BSGs \citep{2014A&A...570L..13C,2023A&A...674A.212D} contradicts the scarcity predicted by the standard single-star evolutionary theory \citep{2024ApJ...967L..39B}, which is referred to as the BSG problem. To resolve this discrepancy, 
several scenarios to create BSGs in addition to the classical post-main sequence objects have been proposed. Namely being stars that experience an extended main-sequence phase due to possessing enlarged convective cores \citep{2011A&A...530A.115B,2021A&A...648A.126M} or due to displaying enhanced rotationally-induced mixing \citep{2013ApJ...764..166D}, post-mass-transfer stars \citep{2019A&A...621A..22F}, stripped-envelope stars \citep{2022A&A...662A..56K}, and binary mergers \citep{2024A&A...682A.169H, 2024ApJ...963L..42M}. In this respect, key to assessing the nature of BSGs and to interpreting their statistics is the thorough understanding of their interior, mass loss, and of their interplay between binary components.

Further contributing to the puzzling demographics of BSGs, several members of the class are believed to have already passed through the phase of red supergiants (RSGs) and are now captured at late stages \citep{2015A&A...575A..60M}. These, so-called post-RSGs, have experienced significant loss of their hydrogen envelope, exposing products from the interior to the surface and surroundings. Spectroscopic tracers of this processed material provide means to identify the evolved status of several of these objects \citep[e.g.][]{2023Galax..11...76K}, although other post-RSGs cease to show abnormal chemistry \citep{2014MNRAS.439L...6G} or a gaseous/dusty circumstellar environment \citep{2009ASPC..412...17O}. The interior of post-RSGs is subject to non-adiabatic processes \citep{2013MNRAS.433.1246S,2024MNRAS.529.4947G}, which are linked to the yet poorly-understood strange-mode instabilities. These have been proposed to be responsible for the launch of episodic mass loss and might lead to a build-up of circumstellar envelopes \citep{2010A&A...513L..11A}. They also require that the stars have experienced copious amounts of mass loss in a short time, during their prior RSG phase \citep{2013MNRAS.433.1246S,2014MNRAS.439L...6G,2024MNRAS.529.4947G}. Moreover, post-RSG BSGs display semi-regular variability and periods of few to several tens of days, which are attributed to opacity-driven oscillations \citep{1997A&A...320..273K,2007A&A...463.1093L}.  These objects are known as \acyg~variables, following the name of the prototype BSG star.

The stellar variability integrates the feedback from the different processes that take place in the interior and atmosphere, serving as a potent indicator of the structural changes that stars undergo at different stages of evolution. From the interior of BSGs, the most commonly studied sources of (quasi-)periodic and stochastic signal include p-/g-modes that are driven by the $\kappa-$mechanism operating in the metal opacity bump \citep[e.g.][]{2006ApJ...650.1111S,2007A&A...463.1093L}, low-frequency waves that are generated by sub-photospheric convection \citep{2009A&A...499..279C}, and internal gravity waves \citep[IGWs;][]{2019NatAs...3..760B,2020A&A...640A..36B}. As soon as the waves reach the surface, they manifest as coherent or stochastic fluctuations in the brightness and variability in the line profiles \citep[e.g.][]{2006A&A...447..325K,2018A&A...612A..40S}. Moreover, pulsations are believed to contribute to line broadening through the effect of macroturbulence \citep{2009A&A...508..409A,2010ApJ...720L.174S,2018MNRAS.476.1234A,2020A&A...640A..36B}. Beyond the interior, rotational modulation is another source of periodic signal that is generated by spots in the surface and/or co-rotating inhomogeneities in the wind \citep{2013A&A...557A.114A,2018MNRAS.476.1234A}. Evidence of the deep interior dynamics propagating into the wind has been demonstrated \citep{2015A&A...581A..75K,2018A&A...614A..91H,2017A&A...602A..32A,2023A&A...677A.176C}, implying an auxiliary role of the pulsations in contributing to the line-driven mass loss. Finally, stellar encounters introduce variability in the case of eclipsing and/or spectroscopic systems, and via indication in the frequency spectra of mixed or tidally-induced modes \citep{2012A&A...542A..88D, 2020MNRAS.497L..19S}.

Probably the most debated type of variability in the frequency spectra of early-type stars is the stochastic low-frequency (SLF) variability. Its first detection in three O-type stars by \cite{2011A&A...533A...4B} has been followed by systematic analyses over extended OB-star samples that confirmed the ubiquity of the signal \citep{2019A&A...621A.135B, 2019NatAs...3..760B, 2020A&A...640A..36B, 2020A&A...639A..81B, 2024ApJ...966..196M, 2024ApJS..275....2S}. Hydrodynamical simulations of core convection and wave propagation have been able to reproduce the observed SLF variability, assigning its origin to IGWs that are excited at the interface between the convective core and the radiative envelope \citep{2013ApJ...772...21R,2019ApJ...876....4E,2023A&A...674A.134R}. Other studies, on the other hand, have demonstrated that core-excited IGWs are unlikely to generate variability consistent to the observations \citep{2019ApJ...886L..15L, 2023NatAs...7.1228A}; a thoroughly discussed alternative to the core-convection scenario is the turbulent motion in the sub-surface convection zones of the stellar envelope \citep{2021ApJ...915..112C,2022ApJ...924L..11S}. The latter excitation mechanism alone, however, is believed to be unfavorable for stars at low-metallicity environments where this form of variability is observed to persist \citep{2024A&A...692A..49B}, as well as for stars in the main sequence that presumably lack sub-surface convection zones  \citep[e.g.][]{2022ApJ...926..221J}. Recent findings  suggest that both core and near-surface convection could shape the discussed signal over different ranges of frequencies in the power spectra of the stars \citep{2024MNRAS.531.1316T}, whereas, the possibility that the inductive role of the two mechanisms relies upon the stellar age and metallicity is too under consideration \citep{2024A&A...692A..49B}. As the discussion on the source of the enigmatic signal continues, refining the knowledge on its underlying physics has important implications for the modeling of stellar evolution, due to the role of the propagating waves in driving the angular momentum transport, rotation and mixing in the interior of the stars \citep{2013ApJ...772...21R,2019ARA&A..57...35A,2020FrASS...7...70B}.

The signatures of the ambiguous processes that govern the nature of BSGs are imprinted into their variability profile. Applying advanced methodology for the exploration of the frequency domain is shown to be effective in identifying different oscillation modes, constraining parameters of the core, and even differentiating between the distinct evolutionary scenarios  \citep{2010Natur.464..259D,2019NatAs...3..760B,2020FrASS...7...70B,2023NatAs...7..913B,2024ApJ...967L..39B,2024A&A...682A.169H}. A particular breakthrough in asteroseismology has been achieved since the launch of NASA’s Transiting Exoplanet Survey Satellite \citep[TESS;][]{2015JATIS...1a4003R}. The mission holds a major advantage over its predecessor monitoring surveys thanks to its all-sky coverage, which enables systematic studies to be undertaken and statistical inferences to be drawn over stellar ensembles. The latter include massive stars at both early phases \citep[e.g.][]{2019ApJ...872L...9P,2020A&A...639A..81B,2021A&A...648A..79K,2022A&A...665A..36G,2022AJ....163..226L} and at evolved (or transition) ones \citep{2021MNRAS.502.5038N, 2019ApJ...878..155D, 2022ApJ...940...27D,2025AJ....169..128S}. These works have been mapping the different manifestations as a function of the stellar parameters, collectively improving our understanding of the stellar structure and evolution. In the same line, we here present a variability study with TESS of southern Galactic BSGs, where we explore associations between their well-determined parameters and the features of the time and frequency domain. Particular attention is paid on the reported \acyg~variables of the sample, probing for signatures of their evolved status on the variability statistics. Additionally, we model the SLF variability of the stars and compare its parameters to those from the study of BSGs in the Large Magellanic Cloud (LMC) \citep{2019NatAs...3..760B}. With the metallicity constrained, we aim to assist in narrowing down the scenarios on the ambiguous origin of the signal. 
The paper is organized as follows: in Sect. \ref{sec:data}, we present the sample of the studied BSGs and the processing of their TESS data, in Sect. \ref{sec:method} we describe the process for extracting the frequencies from the multiperiodic signals and introduce the metrics for describing the time domain, and in Sect. \ref{sec:seds}, we proceed to fit the spectral energy distributions, which enables us to assess the stellar luminosities. We discuss the results of the study in Sect. \ref{sec:results}, and provide concluding remarks in Sect. \ref{sec:summ}.

\section{Sample selection and TESS photometry}
\label{sec:data}

\subsection{Source catalog}
Our explored sample is taken from the catalog of \cite{2010MNRAS.404.1306F}, which consists of Galactic B$-$type supergiants observed in $2004-2005$ with the Fiber-fed Extended Range Optical Spectrograph \citep[FEROS;][]{1999Msngr..95....8K}, a high-resolution spectrograph (${\rm R}\sim48\,000$) installed on the 2.2-m Max Planck Institute (MPI) telescope, at the European Southern Observatory (ESO), La Silla, Chile. The stellar parameters were extracted by fitting the observations with theoretical models, which were generated using the non-local thermodynamic equilibrium (non-LTE) code TLUSTY \citep{1995ApJ...439..875H}. The uncertainties in the log\,$g$ and \logt{}{} measurements were reported as $\pm$0.1 dex and $\pm$0.02 dex, respectively.  

\subsection{TESS time-series data}
 Launched in 2018, TESS \citep{2015JATIS...1a4003R} has been conducting monitoring of $85\%$ of the sky with an angular resolution of $21\arcsec$/pixel, using a wide red/optical passband that spans the wavelength range $600-1\,000$ nm. The field-of-view of TESS covers a sky area of $24^\circ\times96^\circ$ (known as sector), which is observed for two spacecraft orbits, or $\sim$27 days.

We performed cross-matching between the source catalog and the TESS database, removing stars that are not observed by TESS as well as those with poor-quality data (e.g. stars captured at the edge of the detector) that were unsuitable for further investigation. In addition, we excluded stars from the source catalog that are brighter than the TESS magnitude ($T$) limit of 4 mag, a value that has been tested for the ability of the TESS CCD detectors to conserve charge from bright sources \citep{TESSInstrumentHandbook}. Our studied sample consists of 41 BSGs, which we present in Table \ref{tab:sample} of the Appendix, along with their coordinates and spectral classifications from Simbad, their TESS Input Catalog (TIC) number, $T$, and observed sectors. \\

For the systematic analysis of the time series data, we used the \texttt{lightkurve} software package \citep{2018ascl.soft12013L}. Two types of flux were explored; pre-processed and own extracted and systematics-corrected one. 

\subsubsection{\texttt{PDCSAP} flux}
Using the \texttt{lightkurve.search\_lightcurve} routine and the TIC identification number of the objects, we queried the Mikulski Archive for Space Telescopes\footnote{https://mast.stsci.edu/portal/Mashup/Clients/Mast/Portal.html} (MAST) for light curves that have been processed with the dedicated pipeline for the survey (Science Processing Operations Center; SPOC). The light curve files contain data that were taken with a 2-min cadence, and which have been corrected for the systematic trends using the co-trending Basis Vectors (CBVs) method (photometry flagged as \texttt{PDCSAP\_FLUX}; Pre-search Data Conditioning Simple Aperture Photometry). The provided data come along with the pixel mask (``optimal'' aperture) that has been produced by the pipeline for the extraction of the photometric signal.

\subsubsection{Full-frame images (FFIs)}
\label{sec:ffi}

As an adjunct to the study of the pre-processed flux, we explored cutouts from the full-frame images (FFIs), so-called target pixel files (TPFs), using the \texttt{TESScut} service \citep{2019ascl.soft05007B}. The process was performed with the \texttt{lightkurve.search\_tesscut} routine and setting the cutout length to 20$\times$20 pixels, which was enough to encompass the entirety of the target flux and sufficient background signal, for all sample stars. 

We superimposed photometric sources from the \textit{Gaia} Data Release 3 \cite[\textit{Gaia} DR3;][]{2022yCat.1355....0G} database on the TPFs, and located the target star. We then overlaid the SPOC aperture and searched, sector by sector, for cases where adjustment of the aperture was desirable in order to minimize biases related to under-/overcollecting signal. In such an event, we defined a custom mask for manual extraction of the target flux; starting from the pixel containing the \textit{Gaia} target source, we extended the aperture vertically (across the rows) in both directions of the charge blooming, for as long as the ratio of the flux to that of the target pixel remained higher than a threshold value. The process was repeated sideways for the adjacent columns. The threshold was evaluated empirically to include adequate signal of the studied star without excess contribution from the background; it was set to 0.15 for most of the sample stars, and reached down to 0.07. For three stars proximal to a bright source (see Sect. \ref{ssec:flag}), we increased this value up to 0.25. Next, the extracted light curves were corrected against the instrumental noise and the systematics using the \texttt{lightkurve.RegressionCorrector} class. For this process, we defined as background pixels those fainter than a modest threshold ($10^{-4}\sigma$) below the median flux of the TPF. The dimensionality of the background vector space was reduced using principal component analysis, and linear regression was performed to detrend the science data.\\

\noindent The processed data of each above types of fluxes were fit with a low-order polynomial and were detrended. We converted the normalized flux $F_{n}$ to magnitudes using the formula
\begin{equation}
    \Delta m = -2.5 \log_{10}(F_{n})
\end{equation}
The time series of consecutive sectors with the same cadence were then joined into a larger observing window in order to increase the resolution of our frequency analysis and accuracy of time-domain statistics. We display the light curves of selected BSGs in Fig. \ref{fig:lc_0}, and those of the entire sample in the Appendix \ref{app:supp_fig}.

\begin{figure*}
\centering
\includegraphics[width=17cm]{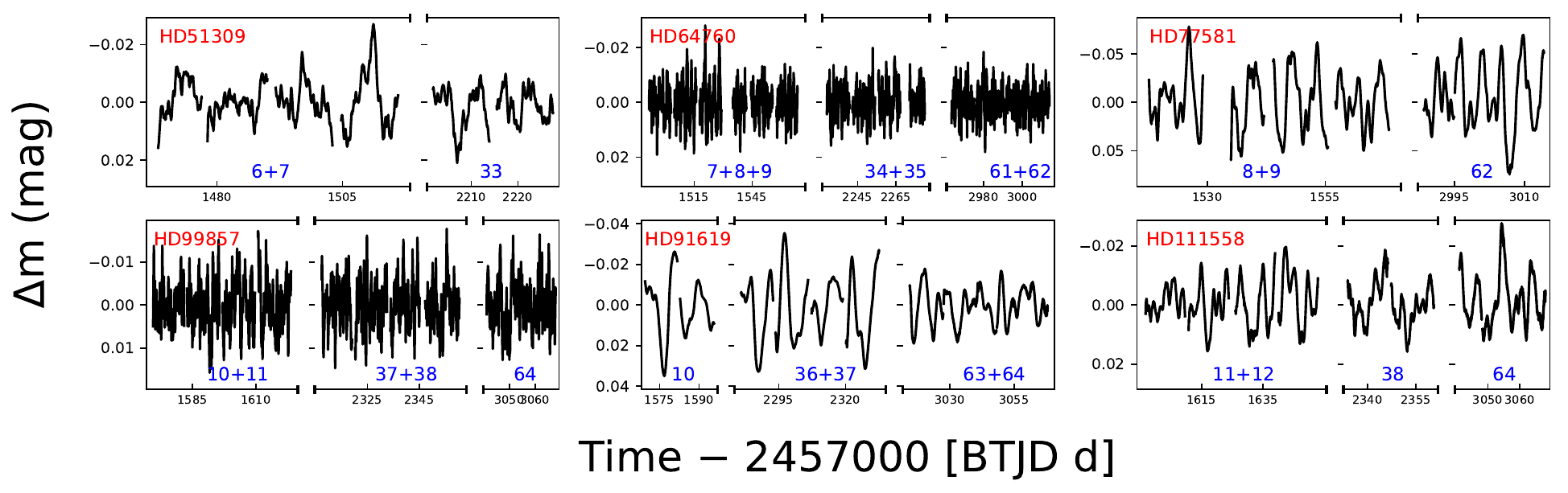}
\caption{\label{fig:lc_0} Time-series photometry from TESS of selected BSGs. Data are detrended, normalized, and converted into units of magnitude. For each star, we show data from different sectors over multiple panels. Consecutive sectors are joined together and their numbers are displayed at the bottom of each panel. The light curves of the entire sample are displayed in the Appendix \ref{app:supp_fig}.}
\end{figure*}

\subsubsection{Contamination in the TESS aperture}
\label{ssec:flag}

The main drawback of TESS is the low spatial resolution, which makes densely populated areas in the sky prone to contamination effects. The degree of contamination for a TESS source can be assessed from the ratio of the light that comes from the target to the total light that is captured within the aperture. This metric is calculated by SPOC and is stored in the header of the light curve files under the keyword \texttt{CROWDSAP}. In cases where a custom/adjusted mask was selected for extracting the signal from the TPFs (Sec. \ref{sec:ffi}), we (re-)assessed the metric as
\begin{equation}
    \texttt{CROWDSAP$^+$} = \frac{f_{*}}{\sum_{k} f_{k}}
\end{equation}
where $f_{*}$ and $f_{k}$ are the fluxes from the target and the $k$-th  source contained in the aperture (including the target), respectively, these being calculated from the TESS magnitudes \citep{2021arXiv210804778P}. A comparison between \texttt{CROWDSAP} and the adjusted values yielded minor differences of $\lesssim0.01$, being only higher for the below listed stars HD\,111973 and HD\,152234.

We provide the contamination metric of the stars in Table \ref{tab:sample}. Values close to 1 indicate that the flux from all secondary sources in the aperture contributes minimally to the target light curve, whereas, the photometry becomes less reliable for values of the metric $\lesssim0.8$ \citep{2021ApJS..254...39G}. The latter case concerns here stars HD\,152234 and HD\,111973. These are followed by HD\,141318 and HD\,79186, with \texttt{CROWDSAP$^+$} values 0.84 and 0.89, respectively. The rest of the sample displays $\texttt{CROWDSAP$^+$}>0.9$, with the vast majority of these stars exceeding 0.95.

Finally, during the inspection of the TPFs (Sect. \ref{sec:ffi}) we flagged three stars, which could be affected by  flux from a bright source that is located outside the photometric aperture. These objects are the above mentioned HD\,111973 and HD\,79186, as well as HD\,94493.

\subsection{Overlapping TESS studies and reported variables}

We checked in TESS studies for overlapping cases and report HD\,51309 to have been studied by \cite{2020A&A...639A..81B} and \cite{2020A&A...640A..36B} with data from Sectors $1-13$. We extend here the variability study to Sector 33 and complement with the time-domain features of the star, measuring also its luminosity from the modeling of multiband photometry.

During a combined cross matching of our stars with the General Catalog of Variable Stars \citep[GCVS;][]{2017ARep...61...80S} and the variability study with Hipparcos by \cite{2009A&A...507.1141L}, nine stars of the sample were identified as \acyg~variables\footnote{We here account for those stars with a solid classification, disregarding those flagged as candidates.}. Moreover, three BSGs are reported in Simbad as double/multiple systems: the above discussed for their crowding effect HD\,152234 and HD\,111973, as well as HD\,77581. The latter is the well studied high-mass X-ray system Vela X-1.

\section{Methodology}
\label{sec:method}

The TESS data of the Galactic BSGs display variability of up to $\sim$0.1 mag in amplitude, which is described as (quasi-)periodic with dominant frequencies that, for several stars, vary across the different observing windows. We identify cases where variability displays irregular patterns that can be also explained as a superposition of cycles with different periods. At a first glance, the signal of several stars with low photometric amplitude resembles that of instrumental noise (the least luminous BSGs; see Sect. \ref{sec:results}), and which typically characterizes the sky objects as ``invariant''.

\subsection{Extraction of frequencies}
\label{ssec:freq}

\begin{figure}
\centering
\includegraphics[width=9cm]{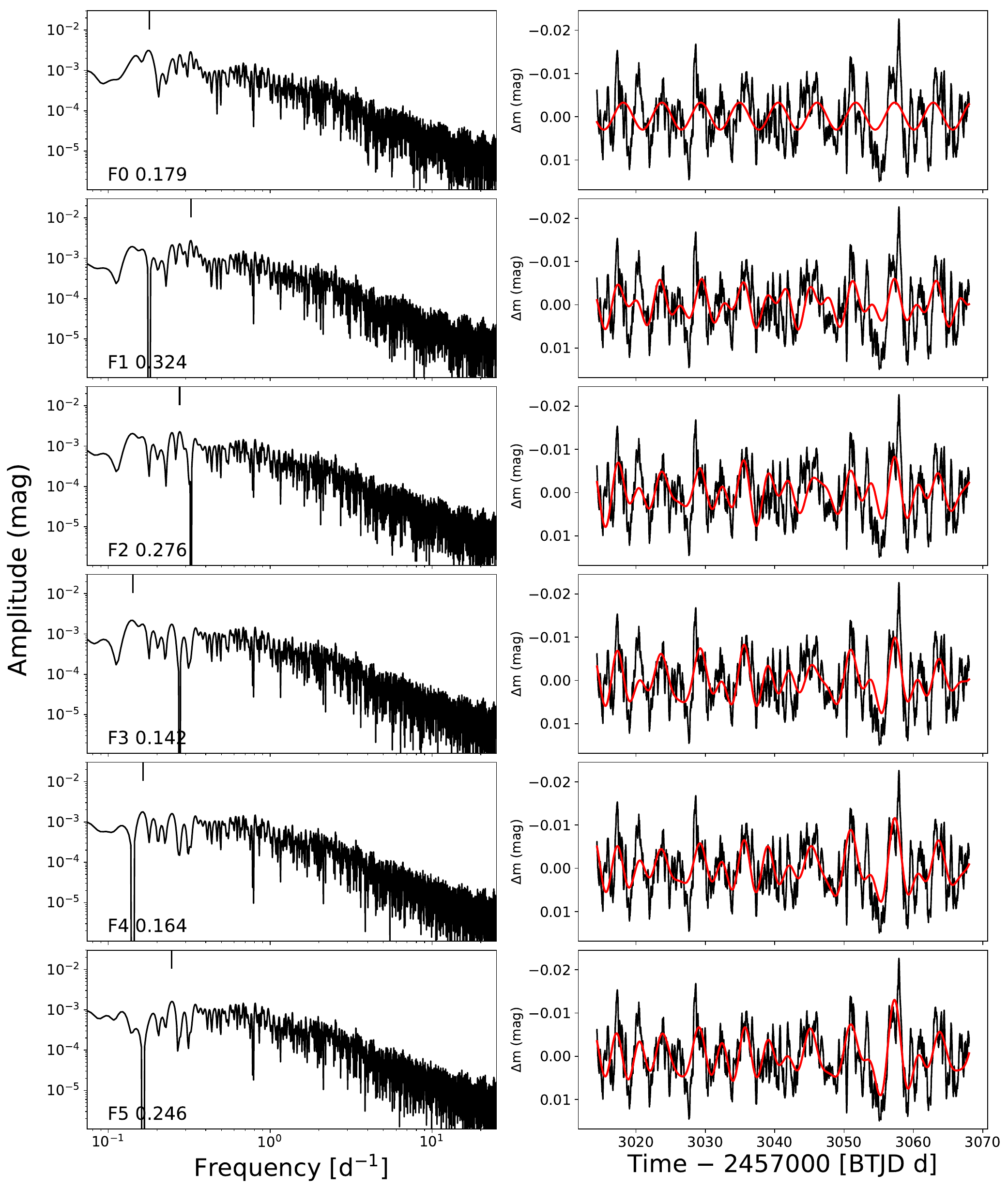}
\caption{\label{prew_HD94493} Iterative pre-whitening for the extraction of frequencies for HD\,94493 (Sec. $63+64$). \textit{Left.} The calculated amplitude spectra (Lomb-Scargle periodograms) shown in logarithmic scale. A vertical line at the top of each panel points to the extracted frequency, which is reported in the lower left. \textit{Right.} The light curve of the star (black points) with superimposed the integrated fit model of sinusoidals (red).} 
\end{figure}

To explore the presence of periodic signals in the TESS data, we made use of the \texttt{lightkurve.Periodogram} class and computed the Lomb-Scargle periodograms (amplitude spectra). We extracted the prominent frequencies via an iterative pre-whitening process; at each iteration, we selected the frequency with the highest amplitude, fit the light curve with the corresponding sinusoidal model and subtracted it. The residuals of the fit were then subjected to a new iteration for identifying the next peak frequency, and the process was repeated until a termination criterion was met (see Sect. \ref{method:rn}). The procedure is demonstrated in Fig. \ref{prew_HD94493} for star HD\,94493 using the stitched light curves from Sectors 63 and 64. 

The frequencies that were extracted from an observing window were categorized into four types using the Rayleigh resolution $1/\Delta T$, where $\Delta T$ is the length of the studied timespan. These types are:
\begin{itemize}
    \item independent frequencies, $f_{i}$. These are significant peaks that are typically identified  early in the extraction process and are physically interpreted in BSGs either as opacity-driven g-/p-modes or as the rotational modulation by an aspherical wind and/or surface spots. 
    \item harmonic frequencies, which were identified as integer multiples of the independent ones.
    \item combinations (algebraic sums) of two previously extracted frequencies.
    \item frequencies that were repeated within the adopted resolution. These are likely to be spurious detections that emerge during the subtraction of the periodic model, and essentially trace the removed signal. The probability of them being unresolved mode orders of period-spacing patterns (and thus having a physical origin) is too considered.    
\end{itemize}
In addition, we considered credible frequencies those being higher than 0.07 d$^{-1}$ in order to resolve a minimum of two cycles per sector. Joining consecutive sectors to form a larger observing window did not lift this constraint, given that the data of the separate sectors were individually normalized, potentially washing out periods longer than $\sim$14 days. The total of the independent frequencies and their parameters, subject of our investigation in Sect. \ref{sec:results}, are tabulated in Appendix \ref{app:supp_ffreq}.

We explored whether the identified frequencies are associated to effects introduced by the stellar rotation, in a similar manner as in \cite{2020A&A...639A..81B}. We calculated the lowest frequency expected from the reported $\upsilon$sin$i$ values by \cite{2010MNRAS.404.1306F} and our inferred radii (see Sect. \ref{sec:seds}),
\begin{equation}
   \nu_{\rm low} = \frac{\upsilon{\rm sin}i}{2 \pi R} \sim 0.02  \frac{\upsilon{\rm sin}i~[\rm km~s^{-1}]}{2\pi(R~ [\rm R_{\odot}])} ~ \rm d^{-1}
\end{equation}
A (conservative) shortest rotational period was also defined using the critical limit for the rotational velocity of a sample star, which is estimated by interpolation of the stellar parameters with the evolutionary models from \cite{2012A&A...537A.146E}. 
Hereafter, we assign the term cROT (candidate rotating variables) to those stars with (independent) frequencies that can be justified as rotationally induced. Respectively, the term nROT points to the rest, nonrotating variables.

\begin{figure}
\includegraphics[width=9cm]{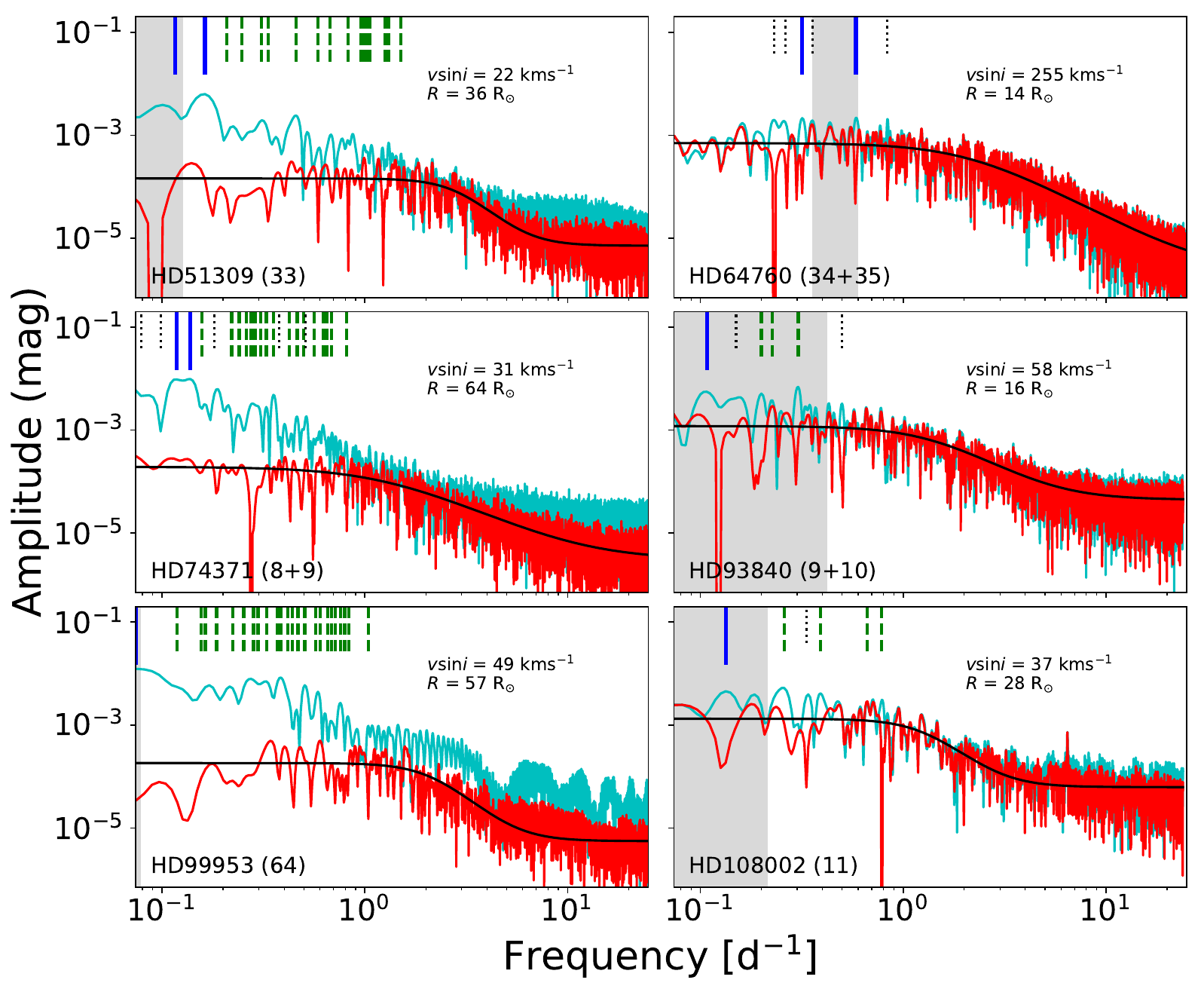}
\caption{\label{fig:ls_0} Frequency spectra of selected BSGs. We show the periodograms calculated at the beginning (cyan) and the end (red) of the pre-whitening process. The vertical lines at the top of each panel point to different types of the identified frequencies; independent (solid), harmonics (dashed), and combinations (dotted). The solid black line corresponds to the best-fit SLF variability model (Eq. \ref{eq:rn}) that we adopt for the star. The grey-shaded region indicates the range of frequencies that can be identified as due to the stellar rotation. We display the data for the entire sample in the Appendix \ref{app:supp_fig}.}
\end{figure}

In Fig. \ref{fig:ls_0}, we display in logarithmic scale the amplitude spectra of selected BSGs at the beginning (cyan) and the end (red) of the iterative process. Among the datasets from different windows per star, we show the one with the best-fit SLF variability model (see below). We mark the different types of frequencies and highlight the range of values (grey-shaded strip) that can be justified as rotationally induced.

\subsection{Stochastic low-frequency variability}
\label{method:rn}

Altogether the amplitude spectra of the Galactic BSGs illustrate a background signal that is indicative of SLF variability. It has the form of a broad excess in the amplitude at low frequencies (red noise), being then followed by a decline of 1$-$2 orders of magnitude and reaching the level that is described as white noise. The stochastically-excited signal is parameterized by a Lorentzian-like profile \citep{2002A&A...394..625S,2011A&A...533A...4B} that is typically employed for describing its properties \citep[e.g.][]{2019A&A...621A.135B,2020A&A...640A..36B,2020ApJ...902...24D,2024ApJS..275....2S}.

At each step of the iterative pre-whitening process, we fit the amplitude (residual) spectra with the function
\begin{equation}
   \label{eq:rn}
      \alpha(v) = \alpha_{w} + \frac{\alpha_{0}}{1 + (2\pi\tau v)^\gamma} \,,
\end{equation}
where \ared{}{} the amplitude of the red noise as $v \to 0$, $\tau$ the characteristic timescale, $\gamma$ the slope of the decay, and \awhi{}{} the white-noise term. The fit of the data was performed in the frequency range 0.07$-$25 d$^{-1}$ using the \texttt{optimize.curve\_fit} routine from the \texttt{scipy} package \citep{2020NatMe..17..261V}. The error estimates were assessed from the covariance matrix. We evaluated the quality of the fit using a standard $\chi^2$ calculation.

We terminated the iterative process as soon as the relative change between the $\chi^2$ values of two successive iterations fell below the tolerance value of 0.1 for five iterations in a row. The signal-to-noise (S/N) values of the prior extracted frequencies were then updated in order to comply with the latest computed model. For describing the SLF variability of a star with multiple observing windows, we adopted the fit model with the lowest $\chi^2$, which we display in logarithmic scale in Fig. \ref{fig:ls_0} (black curve). The 10-base logarithm of the fit values of \ared{}{} and \awhi{}{}, along with the values of $\tau$ and $\gamma$, are listed in Table \ref{tab:cprop}.

\subsection{Metrics of the time domain}

As for a statistical measure of the photometric scatter, we calculated the standard deviation of the magnitude measurements $X_{i}$ 
\begin{equation}
    \sigma = \sqrt{ \sum_{i=1}^{N} {\frac{(X_{i}-\overline{X})^2}{N} } }
\end{equation}
where $\overline{X}$ is the mean magnitude of the light curve and $N$ the number of datapoints. The  alternative and less sensitive to outliers median absolute deviation was also explored; as it delivered almost identical to $\sigma$ results, it is here omitted.

A robust method to assess the independence of successive measurements is the coherency parameter,  $\psi^2$. It characterizes the level of stochasticity in the time series data by evaluating the correlation between the sequential measurements against the  white noise. It is defined on the basis of the zero-crossings $D_{k}$, namely the number of times that the signal crosses the zero level. The non-negative integer $k$ is the order in the time-series differences, such that
\begin{equation}
X_{k,i} = X_{k-1,i} - X_{k-1,i-1}
\end{equation}
with $X_{0,i}$ pointing to the original dataset $X_{i}$. For measuring $D_{k}$, the ``clipped'' data points $Z_{k,i}$ are first generated from $X_{k,i}$ as
\begin{equation}
Z_{k,i} =      
\begin{cases}
1, & \text{if $X_{k,i} \geq 0$} \\
0, & \text{if $X_{k,i} < 0$}
\end{cases}
\end{equation}
wherefrom $D_{k}$ is calculated as the sum of the squared consecutive differences and normalized to their size,
\begin{equation}
D_{k} = \frac{1}{N-k-1}\sum_{i=2}^{N} {(Z_{k,i}-Z_{k,i-1})^2},
\end{equation}
Finally, $\psi^2$ is calculated sufficiently from the first 5 orders \citep{2020MNRAS.497.4843K} as
\begin{equation}
    \psi^2 = \sum_{k=0}^{4} {\frac{(\Delta_{k} - \phi_{k})^2}{\phi_{k}}}
\end{equation}
\noindent where $\Delta_{k}$ denotes the increments of the $k-$order crossings, i.e.
\begin{equation}
\Delta_{k} = 
\begin{cases}
D_{0}, & \text{for $k=0$} \\
D_{k}-D_{k-1}, & \text{otherwise,}
\end{cases}
\end{equation}
\noindent and $\phi_{k}$ being the respective increments of a white-noise signal. The latter values were calculated from a Gaussian-distributed sample of datapoints that was generated over the studied grid of the TESS timestamps.

Finally, we evaluated the asymmetry of the light curves using the moment measure of skewness, which is defined as 
\begin{equation}
     {\rm skw} = \frac{\sum_{i=1}^{N} {(X_{i}-\overline{X})^3}}{N{\sigma}^3} 
\end{equation}
The metric has a positive (negative) value when the distribution of measurements is right (left) skewed, and equals to zero when it is symmetric. \\

The calculated values of the above statistical measures are provided in Table \ref{tab:cprop}. These were taken as the average values of measurements obtained across the different observing windows. Uncertainties are taken as the 1-sigma values.

\section{Spectral energy distributions}
\label{sec:seds}

Our aim is to investigate the association between variability features and fundamental stellar parameters. While \logt{}{} and log\,$g$ for our sample stars have been determined by \cite{2010MNRAS.404.1306F}, their analysis did not provide stellar luminosities.
To retrieve these, we modeled the spectral energy distribution (SED) of the objects.

We constrained the blue part of the SEDs with UBV photometry from \cite{2006yCat.2168....0M} and with data from the Naval Observatory Merged Astrometric Dataset \citep[NOMAD;][]{2004AAS...205.4815Z} that extend to the $R$ band. We included recent photometry from \textit{Gaia} DR3 in $G$, $G_{\rm BP}$ and $G_{\rm RP}$ bands, that in general was found to be in good agreement with the earlier measurements. Near-infrared photometry was taken from 2MASS \citep{2003yCat.2246....0C}, whereas data from the AllWISE catalog \citep{2014yCat.2328....0C} were employed for describing the SEDs up to 22 $\mu$m. To better constrain the flux in the mid-infrared, we also included photometry from the AKARI satellite \citep{2010A&A...514A...1I} and the Midcourse Space Experiment \citep[MSX;][]{2003yCat.5114....0E}. We collected the photometric counterparts by cross-matching between the stellar coordinates and the above catalogs, using a search radius of 2$\arcsec$ for the optical and 2MASS studies and 5$\arcsec$ for the mid-infrared ones that have a lower spatial resolution. The conversion between magnitudes and fluxes was performed based on the passband zeropoints and effective wavelengths, which are available by the SVO Filter Profile Service\footnote{http://svo2.cab.inta-csic.es/theory/fps/}.

We fitted the multiband photometry within a standard optimization procedure using the public available grids of synthetic spectra generated with the TLUSTY code \citep[OSTAR2002 \& BSTAR2006;][]{2007ApJS..169...83L}. We selected model SEDs with solar abundances. As the lower limit of the BSTAR2006 grid is 15\,000 K, we fit the data of our cooler seven stars with theoretical SEDs from \cite{2011MNRAS.413.1515H}, which are generated with the ATLAS9 code.

The flux $f_{\lambda}$ at wavelength $\lambda$ that is received from a star with radius $R$ at distance $D$ and reddened with extinction $A(\lambda)$, is given as 
\begin{equation}
    \label{eq:sed}
    f_{\lambda}=C^{2} F_{\lambda} 10^{-0.4 A(\lambda)} \\
\end{equation}
where $F_{\lambda}$ is the flux emerging from the stellar surface and $C=R/D$. We ran a Levenberg-Marquardt method to minimize the residuals between Eq. (\ref{eq:sed}) and the photometry, setting free the scaling factor $C$ and the visual extinction $A_{V}$ \cite[e.g. as in][]{2015A&A...582A..42K,2022MNRAS.511.4360K}. Relative to the latter, the extinction at $\lambda$ was expressed using the law of \cite{1989ApJ...345..245C} assuming $R_{V}=3.1$. The flux per unit surface area, $F_{\lambda} \equiv F_{\lambda}(T_{\rm eff},{\rm log}\,g)$, was determined for each star by interpolating within the grid of synthetic models. 

The photometry and best-fit models for selected stars are displayed in Fig. \ref{fig:seds_0}, and for the entire sample in Appendix \ref{app:supp_fig}. For several cases, a weak offset between the model flux and the observations in the infrared implies excess due to stellar winds. In Table \ref{tab:sed_prop}, we list the 10-base logarithms of $C$ (expressed in units of \rsun/pc) that are used for determining the stellar radii, thus luminosities, from the distances to the objects. In this work, the latter values were taken as the photogeometric estimates from \textit{Gaia} DR3  \citep{2021AJ....161..147B} and are listed in Table \ref{tab:sed_prop}, along with the resulting \logl{}{} values. 

\begin{figure}
\includegraphics[width=\linewidth]{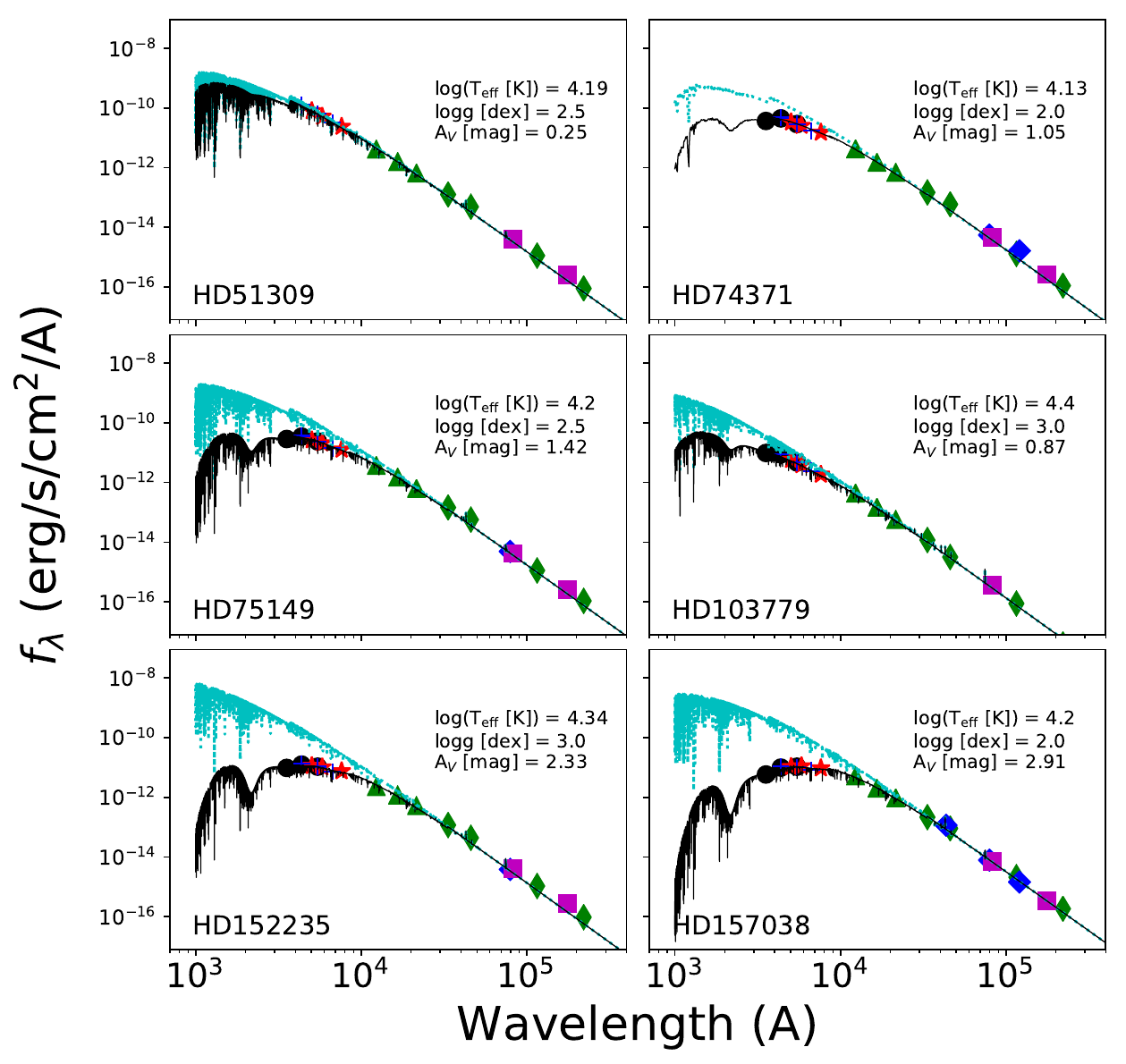}
\caption{\label{fig:seds_0} Spectral energy distributions of selected BSGs. Data were taken from \cite{2006yCat.2168....0M} (black circles), NOMAD (blue crosses), \textit{Gaia} DR3 (red asterisks), 2MASS (green triangles), WISE (green rhombi), AKARI (magenta squares), and MSX (blue diamonds). The best-fit TLUSTY model (ATLAS9 for HD\,74371) is overplotted, unreddened (cyan line) and reddened according to the inferred $A_{\rm V}$ (black line). In each panel, we display the stellar designation (lower left) and parameters (upper right). The modeled SEDs of the entire sample are displayed in the Appendix \ref{app:supp_fig}.}
\end{figure}

\section{Results and discussion}
\label{sec:results}

In the current section, we present and discuss associations between the physical and evolutionary properties of the stars and the features extracted from the analysis of variability.

\begin{figure}
\centering
\includegraphics[width=\linewidth]{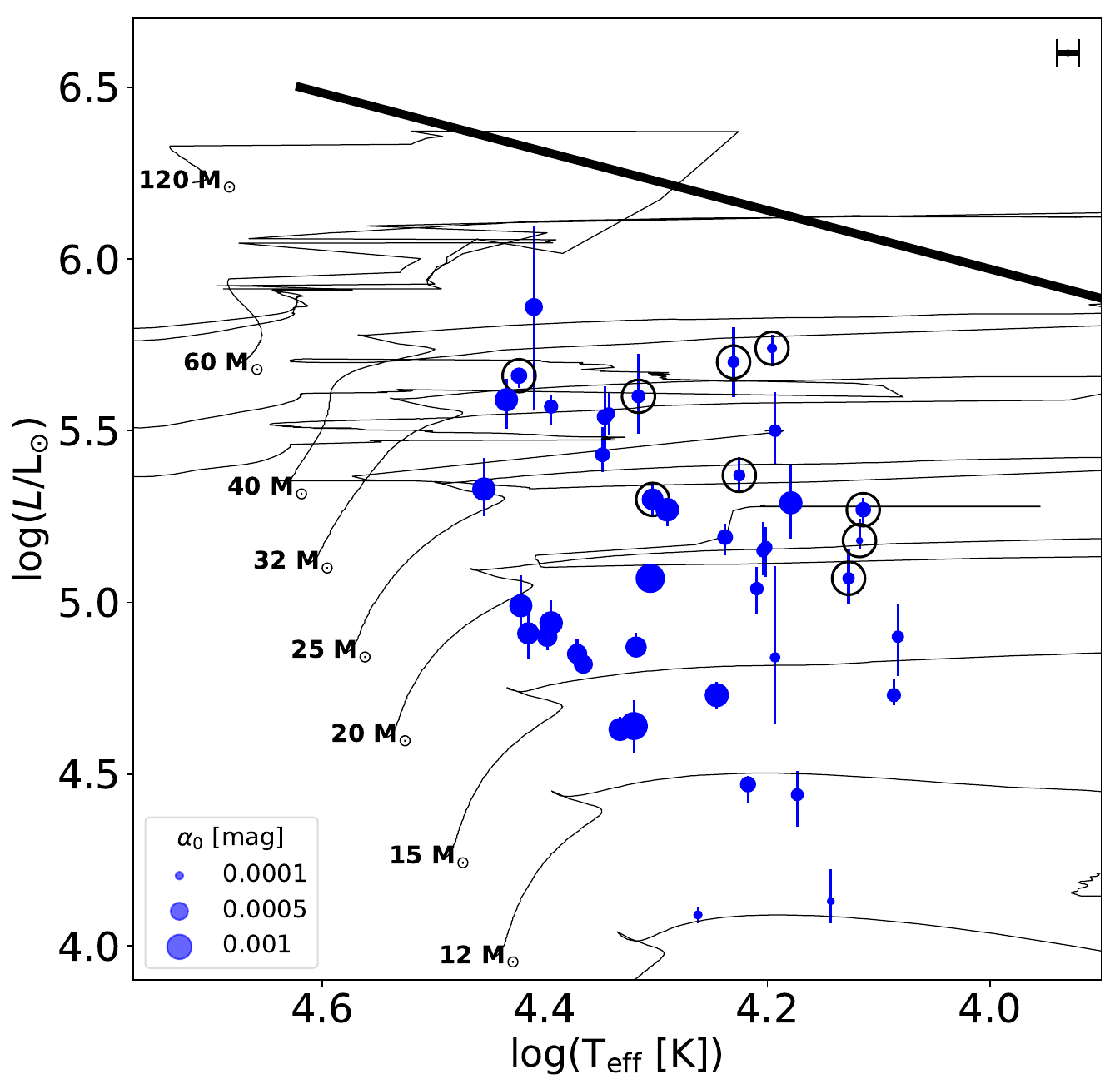}
\caption{\label{fig:hr} Hertzsprung–Russell diagram for the evolution of massive stars. Theoretical tracks at solar metallicity are taken from \cite{2012A&A...537A.146E}, for stars rotating initially at 40\% of their critical velocity. The size of the markers is proportional to the amplitude of the SLF variability (Sect. \ref{ssec:slfv}). The variables of the \acyg~class are indicated by outer circles, and the thick line represents the Humphreys-Davidson limit. The uncertainty in the spectroscopic temperature is illustrated by the errorbar on the upper right.}

\end{figure}

With the stellar luminosities measured, we locate our studied objects in the Hertzsprung–Russell diagram of Fig. \ref{fig:hr}, and superimpose them on models of stellar evolution at solar metallicity \citep{2012A&A...537A.146E}. The tracks assume an initial rotational velocity at 40\% of the critical speed. As can be seen, the sample stars span evolutionary tracks beyond the end of the main-sequence phase in the range $M_{\rm ini}=9-40$ M$_{\odot}$. In the same plot, we indicate the reported \acyg~variables using large outer circles, which occupy the upper right region of the evolutionary diagram. Being in a good agreement with their definition as post-RSGs, they span tracks that suggest evolution bluewards ($5.0\lesssim{\rm log}(L/$\lsun$)\lesssim5.8$). 

As for an inherently spectroscopic assessment and alternative to the SED-based luminosity, we additionally calculated the Eddington factor $\Gamma_{e}$ of the studied objects (Table \ref{tab:sed_prop}). Being function of the so-called spectroscopic luminosity \citep{2014A&A...564A..52L}, $\Gamma_{e}$ is defined as 
\begin{equation}
    \Gamma_{e}\equiv\frac{\kappa_{e}~L}{4\pi~GMc}=\frac{\sigma_{B}~\kappa_{e}}{c}\frac{T_{\rm eff}^4}{g}
\end{equation}
where $\sigma_{B}$ is the Stefan–Boltzmann constant, c the speed of light, and $\kappa_{e}=0.4$ cm$^2$g$^{-1}$ the opacity, here assuming only due to photon scattering by free electrons.

\subsection{Correlations in the time domain}
\label{ssec:tdcor}

\begin{figure*}
\centering
\includegraphics[width=11cm]{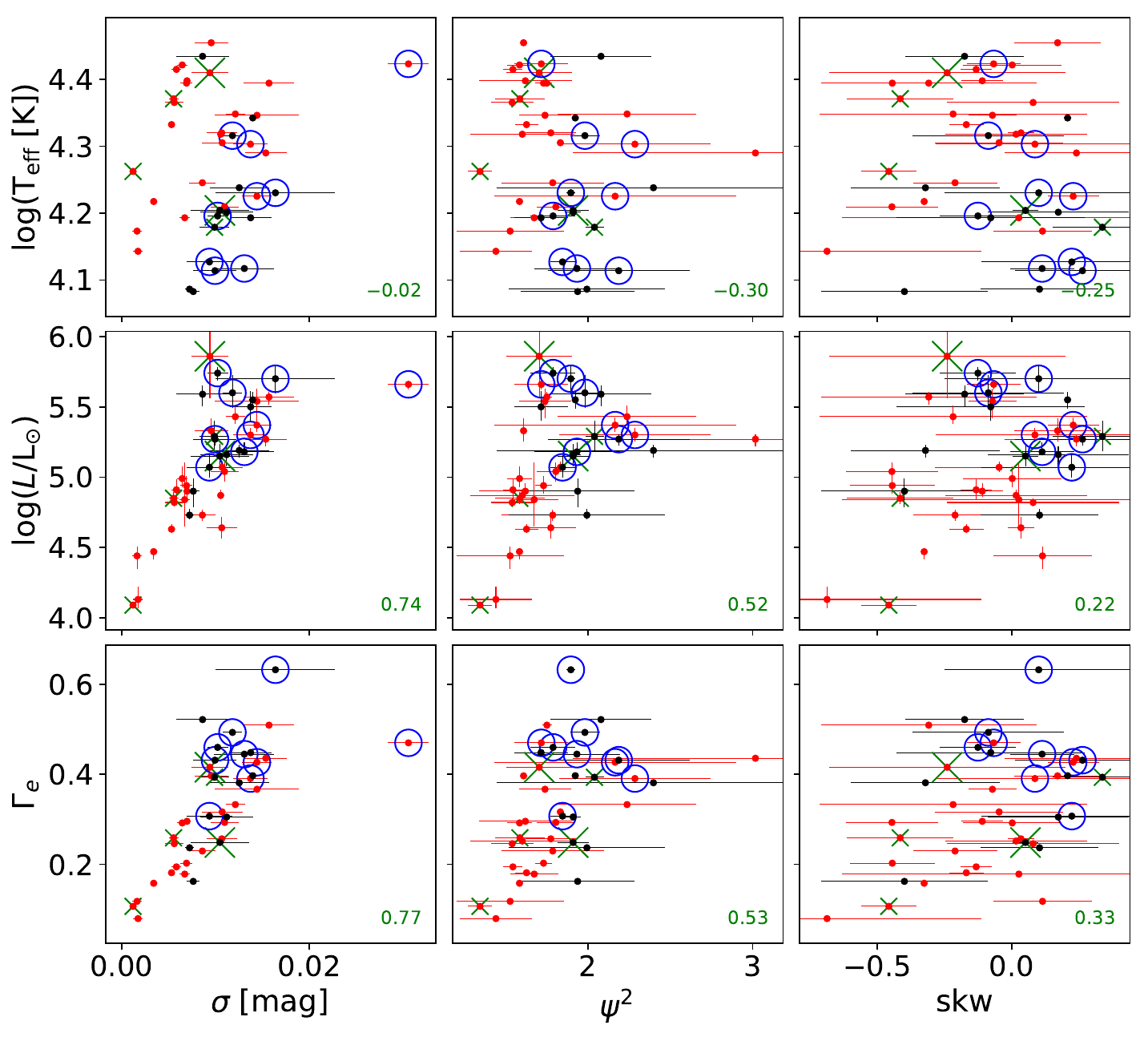}
\caption{\label{fig:sc_par} Plots of the stellar parameters of BSGs against their time-domain statistics. Red points indicate the cROT variables, and black points the nROT ones. The outer circles indicate the $\alpha$ Cygni variables. We mark stars susceptible to TESS contamination with \texttt{CROWDSAP$^+$} values lower than 0.8 (``X'' symbols) and within $0.8-0.9$ (``x'' symbols). To the latter stars, we added HD\,94493 due to its proximity to a bright source located outside the photometric aperture.}
\end{figure*}

In Fig. \ref{fig:sc_par}, we present the plots of \logt{}{}, \logl{}{}, and \gammae{}{} against the calculated metrics of the time domain. We distinguish between the cROT (red markers) and nROT (black markers) variables, and indicate the reported variables of the \acyg~class as in Fig. \ref{fig:hr}. Moreover, we indicate stars susceptible to contamination in their light curves, showing \texttt{CROWDSAP$^+$} values lower than 0.8 (``X'' symbols) and within $0.8-0.9$ (``x'' symbols). To the latter set of stars, we added HD\,94493 for a possible such effect caused by a bright star external to the TESS aperture (Sect. \ref{ssec:flag}). To assess whether the relationship between two parameters is monotonic, we show at the lower right of each panel the Spearman's rank correlation coefficient $r_{s}$. Outliers were excluded from this calculation by masking values whose absolute difference from the median exceeded five times the median absolute deviation.

The strongest correlation found in this study is the positive trend between $\sigma$ and log\,($L$/\lsun) ($r_{s}=0.74, p<0.0001$), which  decodes into a more tight relationship with $\Gamma_{e}$ ($r_{s}=0.77, p<0.0001$). Such a correlation is already reported by the early variability studies of BSGs \citep{1972A&A....20..437M,1980A&A....90..311M}, and  mirrors, to a certain extent\footnote{Particularly for stars that do not display significant isolated frequencies.}, the corresponding trend that is reported between the luminosity and the amplitude of SLF variability \citep{2019A&A...621A.135B, 2020A&A...640A..36B}. Here, the extreme outlier star ($\sigma\sim0.03$) is the \acyg~variable HD\,77581 (Fig. \ref{fig:lc_0}), which as previously mentioned is a member of the eclipsing binary Vela X-1; the star is long known to possess an enigmatic light curve that is modulated by effects such as, gravitational distortion \citep{1973ApJ...184L.121J}, gas streaming \citep{1991ApJ...371..684B}, and non-radial oscillations of the BSG being tidally induced by the companion pulsar \citep{2003A&A...401..313Q}.

Positive correlation is also seen between $\psi^2$ and \logl{}{}  ($r_{s}=0.52, p<0.001$), indicating that the more luminous (and higher amplitude) variables possess smooth light curves with a higher degree of coherency compared to less luminous ones \citep[e.g., also][]{2022A&A...668A.134B}. Moreover, the highest $\psi^2$ values are found to largely fluctuate across different TESS sectors. With several of these cases being flagged as cROT variables, these stars could expose an ephemeral modulation induced by rotation that emerges on top of their pulsational activity. On the other hand, the less luminous BSGs (\logl{<}{5}) are prone to a noise-like photometric behavior, being further identified in their vast majority as cROT stars. Their variability may therefore be explained by the presence of surface spots and/or co-rotating inhomogeneities in the wind \citep{2004MNRAS.351..552M,2013A&A...557A.114A,2015MNRAS.451.1445B,2020A&A...639A..81B}, with the latter features growing with increasing mass loss \citep[thus with increasing luminosity;][]{2021A&A...647A..28K}. The positive correlation between  $\Gamma_{e}$ and the skewness ($r_{s}=0.33, p=0.04$; Fig. \ref{fig:sc_par}) could credit the latter scenario;  winds structured by line-driven instabilities are linked to short-term brightening of the stars, producing thus time series that are negatively skewed \citep{2021A&A...648A..79K}. 

Several \acyg~variables within the upper luminous sample appear to be less bound to the above trends. Moreover, other luminous  stars\footnote{HD\,92964, HD\,108002, HD\,109867 ($\psi^2=2.2\pm0.4$), HD\,111973, HD\,115842, HD\,152235, and HD\,154090.} (\logl{\gtrsim}{5.1}), as well as the less luminous  nROT variable HD\,105071 (with $\psi^2=2.0\pm0.5$; Fig. \ref{fig:sc_par}), exhibit the \acyg~phenomenon in their light curves, and are marked as ``candidate'' \acyg~variables in the Hipparcos study of \cite{2009A&A...507.1141L}. Collectively, the irregularity in the photometric data of these objects follows the  realization that they are not homogeneous with respect to the excitation mechanisms of their pulsations, these including oscillatory convection modes and radial strange modes \citep{2009A&A...498..273G,2011MNRAS.412.1814S,2013MNRAS.433.1246S}. On top of this activity, a structured and/or variable wind has been reported in stars of the class \citep{2014A&A...566A.125C,2015A&A...581A..75K,2023A&A...677A.176C}.

\subsection{Correlations with the independent frequencies}

\begin{figure}
\centering
\includegraphics[width=9cm]{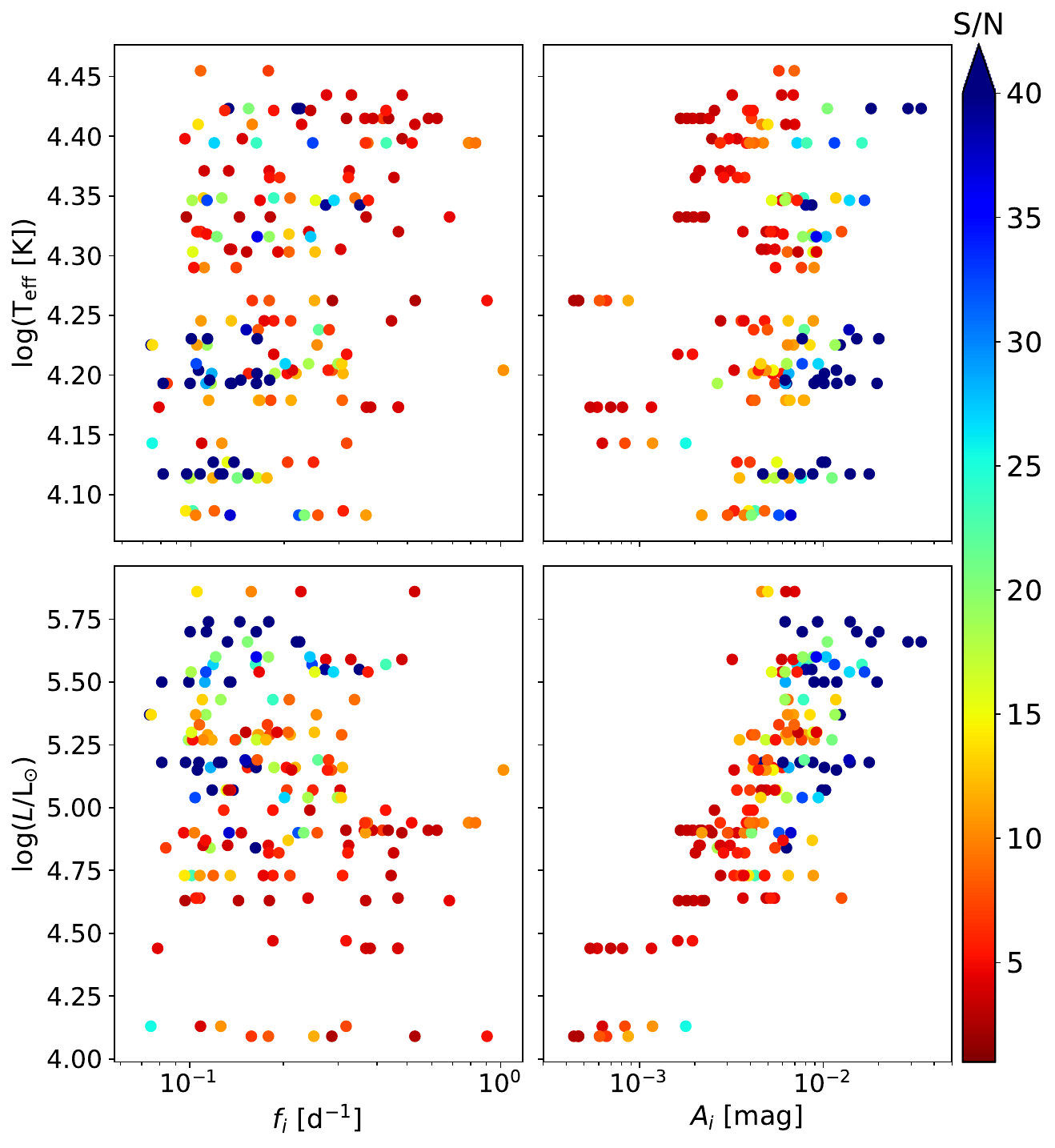}
\caption{\label{fig:sc_freq} Plots of stellar parameters against the frequencies $f_i$ and their amplitudes $A_i$, which are extracted from the periodograms of all observing windows. The data are color coded with respect to their S/N.}
\end{figure}

In Fig. \ref{fig:sc_freq}, we display the total of the independent frequencies $f_{i}$ extracted and their amplitudes $A_{i}$ as function of the \logt{}{} (upper panel) and \logl{}{} (lower panel). The data are color coded as function of the S/N. The number of the frequencies $f_{i}$ per observing window does not exceed three. Accordingly, the number of ``individual'' frequencies per star (under the Rayleigh resolution) that were extracted from the different windows is less than five. Nonetheless, no clear association is found between these metrics and the stellar parameters. We note a disposition of the hotter and less luminous BSGs toward higher frequencies. With the latter showing the lowest S/N values, such an observation is apparently sensitive to the extraction process followed (e.g. the termination criterion for the prewhitening process and the noise or significance level adopted). The distribution of $A_{i}$ mirrors that of $\sigma$ (Fig. \ref{fig:sc_par}), linking the photometric dispersion of stars with \logl{\gtrsim}{5} to significant (quasi-)periodic cycles, whereas moving to the less luminous BSGs, variability is rather described as blends of weak low-amplitude signals. 

\begin{figure}
\includegraphics[width=9cm]{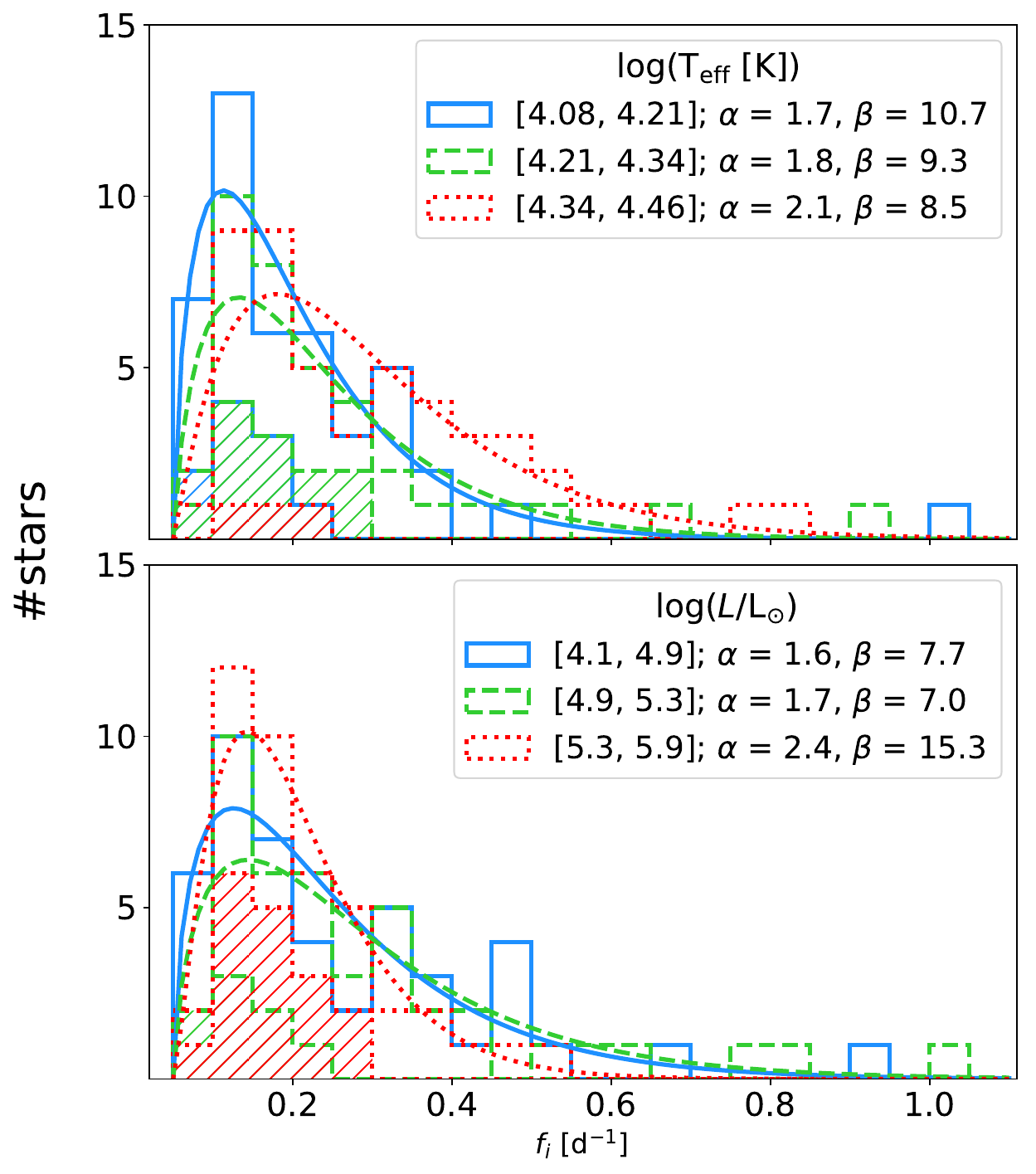}
\caption{\label{fig:hist} Histograms of the individual BSG frequencies as function of \logt{}{} (upper panel) and \logl{}{} (lower panel). In each plot, the sample is split into three groups based on their parameter values (legend). The histograms for the \acyg~variables are indicated with a dashed surface.}
\end{figure}

A more comprehensive view of the latter plot is gained in the histograms of Fig. \ref{fig:hist}, which show the distribution of $f_{i}$ as function of \logt{}{} (upper panel) and \logl{}{} (lower panel). For building the histograms, we employed the individual frequencies per star, i.e. we grouped the frequencies extracted from the different windows together and bin averaged. The size of the bin was chosen to be 0.05 d$^{-1}$ and approximate to the Rayleigh resolution. Then, for each of the two stellar parameters we split the sample by the 33rd and 66th percentile, forming three nearly equal size groups. Their histograms are illustrated with different color/shape; solid blue ($p_{0}$ to $p_{33}$), dashed green ($p_{33}$ to $p_{66}$), and dotted red ($p_{66}$ to $p_{100}$). The interval parameter values are given in the legend. As the data display positive skewness, we fit the histograms to gamma distributions following the two-variable parameterization, with a probability density function of
\begin{equation}
    f(x,\alpha,\beta) = \frac{\beta^{a}x^{a-1}e^{-\beta x}}{\Gamma(\alpha)}
\end{equation}
where $\Gamma (\alpha)$ is the gamma function. The calculated fit models are displayed in Fig. \ref{fig:hist} as thick curves; their parameters are provided in the legend. 

The inspection of the grouped data as function of the stellar temperature (Fig. \ref{fig:hist}, upper panel) reveals that the hot BSGs of the sample with \logt{\geq}{4.34} display shift of their frequencies to higher values. The high-frequency  tail recedes with decreasing temperature such that, in the cooler groups, frequencies become heavily centered around $\sim$0.13 d$^{-1}$. This effect, however, is produced by the higher fraction of the contained \acyg~variables; the distribution of the latter is shown on the plots as hatched histograms. Their identified frequencies are dispersed modestly around the latter value and confined below 0.3 d$^{-1}$. Moving into the statistics on the stellar luminosity (Fig. \ref{fig:hist}, lower panel), one can tell that the aforementioned tail identifies stars of the low- and moderate-luminous groups, which share similar properties.

Summing up, the periodograms differentiate between distinct groups of stars that fall under the BSG umbrella: moderate/high-luminous stars showing frequencies at the lower range, the subgroup of \acyg~stars with $f_{i}<0.3$ d$^{-1}$, and the younger and less massive BSGs with frequencies shifted  to higher values. As the latter group refers to more compact objects per se, we speculate that this effect is caused by the rotational modulation acting over shorter periods and/or to the effects that the Coriolis force exerts on the excited modes, although the latter mechanism has been mostly observed in B-type stars in the main sequence \citep[e.g.][]{2014A&A...569A..18S}.

\subsection{Correlations with the stochastic low-frequency variability}
\label{ssec:slfv}

\begin{figure*}
\centering
\includegraphics[width=11cm]{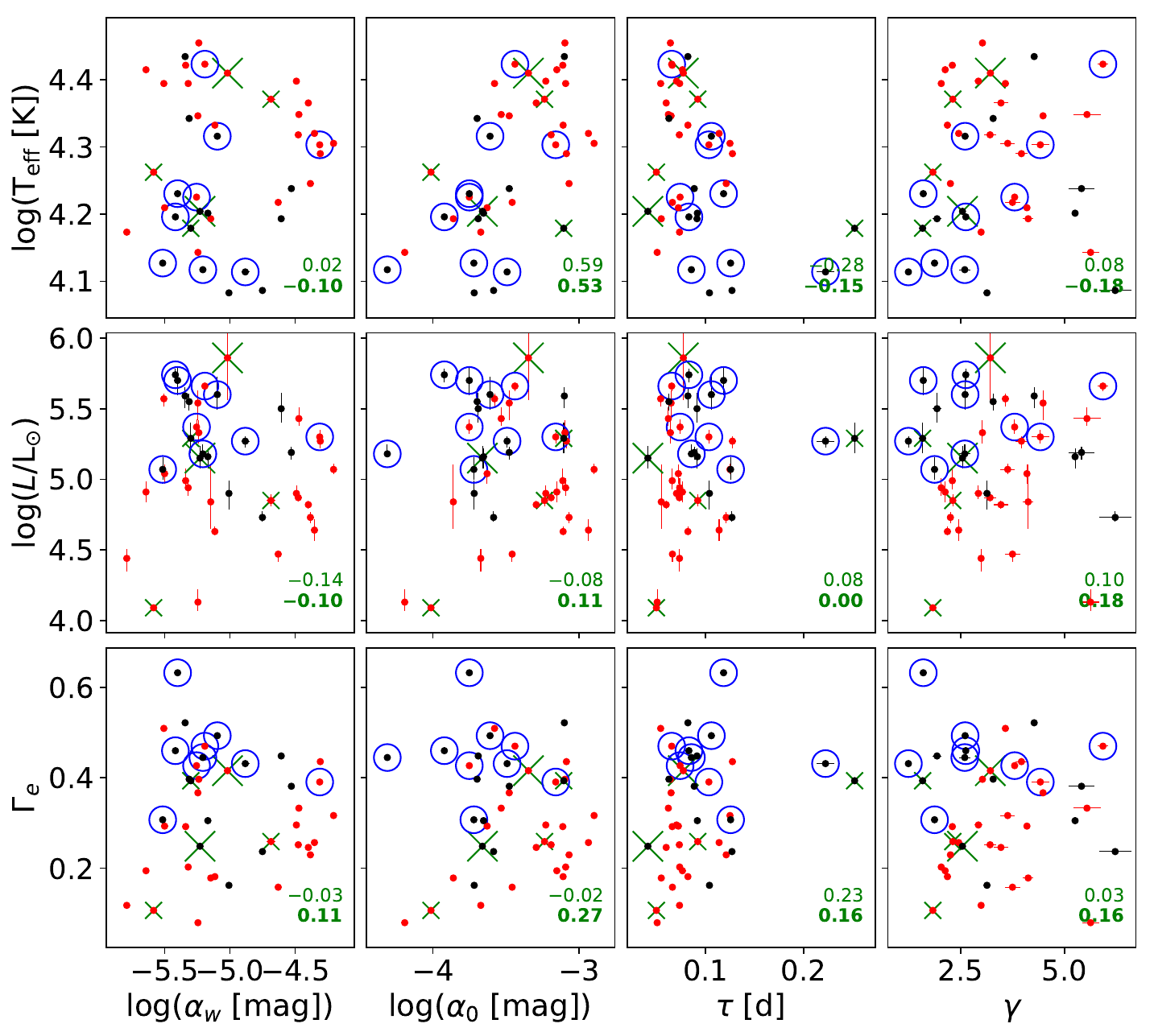}
\caption{\label{fig:sc_rn} Same as in Fig. \ref{fig:sc_par}, for the parameters log\,\awhi{}{}, log\,\ared{}{}, $\tau$, $\gamma$, of the SLF variability models. The bold font value shows Spearman's rank correlation coefficient for a set of parameters, having excluded the \acyg~variables.}
\end{figure*}

In a similar way to Fig. \ref{fig:sc_par}, we show plots of the stellar parameters against the parameters of the SLF variability in Fig. \ref{fig:sc_rn}. To discuss the dependence of the latter on the evolutionary stage of the stars, we re-calculated the Spearman's rank correlation coefficient for each parameter set having excluded the \acyg~stars (bold font value).

A significant positive correlation is seen between the red-noise amplitude and \logt{}{} ($r_{s}=0.59, p<0.0001$) that persists with excluded (and within) the group of \acyg~variables ($r_{s}=0.53, p=0.002$). This information is illustrated on the Hertzsprung–Russell diagram (Fig. \ref{fig:hr}) by setting the marker size to being proportional to \ared{}{}.
Our results suggest that the background signal is more pronounced at earlier stages of evolution and drops as the stars age. Interestingly, an inverse association between these two parameters has been reported by studies on early-type stars spanning the upper main sequence \citep{2020A&A...640A..36B,2024ApJS..275....2S}. Complementing with the present findings over a more evolved sample lets us speculate that SLF variability is amplified at or shortly after the end of the main-sequence phase. 

From Fig. \ref{fig:sc_rn}, it can also be seen that the characteristic timescale $\tau$ is negatively correlated with \logt{}{} ($r_{s}=-0.28, p=0.08$), which is in line with the reported decrease in the corresponding characteristic frequency of the signal toward more evolved stars \citep{2013MNRAS.430.1736S,2020A&A...640A..36B,2023ApJ...954..171V,2024ApJS..275....2S}. The nonsignificant positive trend between $\tau$ and $\Gamma_{e}$ ($r_{s}=0.23$) then mirrors the inverse association of the former  with log\,$g$, whereas no association is seen between $\tau$ and \logl{}{}. The slope $\gamma$ of the red noise, on the other hand, is shown to be most related with \logl{}{}, though nonsignificantly ($r_{s}=0.18$, when excluding the \acyg~stars\footnote{The outlier nROT star with $\gamma\sim6$ is HD\,105071, which as earlier mentioned is a candidate \acyg~variable. Given also its large $\psi^2$ value, we suggest that the distance to the target may be underestimated.}). Weak (or no) correlations are observed between the stellar parameters and the level \awhi{}{} of the white noise.

Studies of young massive stars have highlighted the positive trend between the amplitude of SLF variability and luminosity \citep[e.g.][]{2020A&A...640A..36B,2024ApJS..275....2S}, which are interpreted as due to the larger convective velocities with increasing stellar mass and luminosity of the core \citep[e.g.][]{2013MNRAS.430.1736S,2023A&A...674A.134R,2023NatAs...7.1228A}. This trend is discussed to be insensitive to the parameter of metallicity, which serves as an evidence of the core-excited nature of the background signal, especially for stars at young stages \citep{2019NatAs...3..760B,2024A&A...692A..49B}. At first glance, the analysis of the current, post-main-sequence, sample suggests that log\,\ared{}{} and the stellar (and spectroscopic) luminosities are unrelated (Fig. \ref{fig:sc_rn}). A positive, yet nonsignificant, trend between these parameters is shown when one excludes the \acyg~variables.

We compare the questionable latter trend to the relevant findings from \cite{2019NatAs...3..760B} in the LMC. Using TESS data, these authors extracted the parameters of the SLF variability from 53 BSGs with  spectral type earlier than B4. For our purpose we calculated, and provide in Table \ref{tab:sed_prop}, the absolute magnitudes in $G$-band of our sample stars from
\begin{equation}
    M_{G}=G-5{\rm log}_{10}D+5-A_{G}
\end{equation}
using the values of distance $D$ from \textit{Gaia} (Table \ref{tab:sed_prop}). The $G$-band extinction was estimated from our SED-estimated extinction in the visual considering that $A_{G}\simeq0.77 A_{V}$ \citep{2018MNRAS.481.4093S}. 

\begin{figure}
\centering
\includegraphics[width=\linewidth]{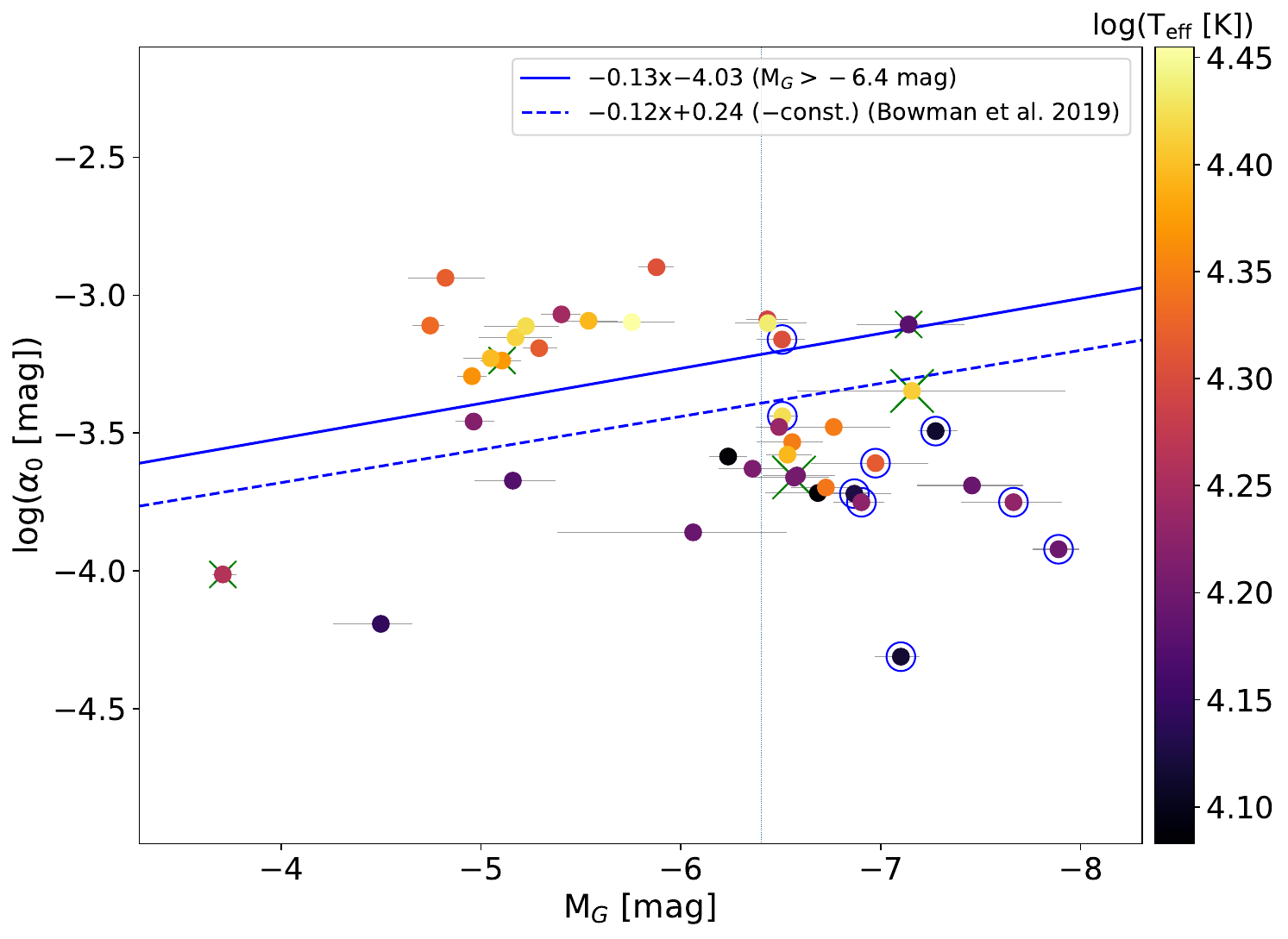}
\caption{\label{fig:LMC} Absolute $G-$band magnitudes vs log\,\ared{}{}. The markers are color coded as function of \logt{}{}. The designation for the \acyg~variables and for stars susceptible to TESS contamination follows that in Fig. \ref{fig:sc_par}. We show the linear fit to BSGs with $M_{G}$ fainter than -6.4 mag (thick solid line). The fit (+offset) to the respective parameters of BSGs in the LMC is displayed \citep[dashed line;][]{2019NatAs...3..760B}.}
\end{figure}

In Fig. \ref{fig:LMC}, we display $M_{G}$ against log\,\ared{}{}, color coding the markers as function of \logt{}{}. The variables of the \acyg~class (with large outer circles) appear confined below $-$6.4 mag (vertical thin line). Considering that possible unidentified \acyg~stars would occupy the same area on the plot (see also Sect. \ref{ssec:tdcor}), we focus on the 25 BSGs that are fainter than this threshold. As expected from our above discussion, a nonsignificant weak (negative) trend is measured between the amplitude of the SLF variability with decreasing $M_{G}$  ($r_{s}=-0.08$). We depict the linear fit to the truncated sample using a solid thick line.

An important remark in the study by \cite{2019NatAs...3..760B} is the absence of coherent frequencies in the LMC sample, which led the authors to proceed in modeling the SLF variability without pre-whitening their data. As the latter process wears down the amplitude spectra (e.g., see Fig. \ref{fig:ls_0}), the relevant properties are expected to be systematically offset compared to when the original data are modeled instead\footnote{For testing purposes, we explored our sample  using the original amplitude spectra and confirm the effect.}. Such a speculation is also evident within the study of \cite{2019NatAs...3..760B} where a sample of eclipsing stars was explored additional to their BSGs, nevertheless following  pre-whitening of their signal peaks. When comparing between those two samples, the measured values for the red-noise level are shown to be systematically offset (their Fig. 4). We display the fit model from the discussed study in our Fig. \ref{fig:LMC} as a dashed thick line. For the sake of comparison, an arbitrary constant is added to it in order to match the mean level of our data\footnote{This constant accounts also for the conversion among units.}. 

Similar trends are found between the study of \cite{2019NatAs...3..760B} and the current work when confining ourselves to lower luminosities  (Fig. \ref{fig:LMC}). In this regime, one may speculate about a metallicity-independent mechanism
for driving the SLF variability, such as core convection. Furthermore, as seen in Fig. \ref{fig:sc_rn}, the vast majority of variables with \logl{\lesssim}{5} display $2\lesssim\gamma\lesssim4$, which match, in general, the power-law exponents of the model spectra generated by core-excited IGWs \citep{2019ApJ...876....4E}. We do stress that a larger statistical sample is essential for establishing the present result, given the sparsity of our sample toward the low-luminosity end where also the TESS data of HD\,141318 are susceptible to contaminating flux ($\texttt{CROWDSAP$^+$}=0.842$). The results from the current study make also less plausible the scenario that SLF variability is generated by sub-surface convection zones driven by the iron opacity peak. The convective flux in such zones relative to the total stellar flux is expected to increase with decreasing \logt{}{} and increasing luminosity \citep{2021ApJ...915..112C,2024ApJS..275....2S}, which does not comply with the current picture for the SLF variability amplitude; instead, we observe a decline in \ared{}{} toward the upper cool Hertzsprung–Russell diagram (see Fig. \ref{fig:hr}).

From Fig. \ref{fig:sc_rn}, it is also evident that ``within'' the \acyg~group no trend is evident between log\,\ared{}{} and our both luminosity assessments (accordingly, between log\,\ared{}{} and $M_{G}$, for $M_{G}\lesssim-6.4$; Fig. \ref{fig:LMC}). Rather, these stars are en masse predisposed to lower SLF variability amplitudes. As the stellar age appears in this study to be more influential for the latter parameter compared to the mass, we attribute this effect to the advanced evolutionary stage of \acyg~stars. In the scenario that the SLF variability is core-excited, the suppressed signal of \acyg~variables could mirror stars with undersized cores, e.g. due to efficient mixing of the core material into the stellar envelope at the earlier phases. An alternative explanation would point to radiative damping of the low-frequency modes when these propagate within a density stratified envelope, with the effect becoming more prominent with the stellar age \citep[e.g.][]{2023ApJ...954..171V}. Finally, the positive association that is seen in the early stages between the red-noise amplitude and the stellar mass \citep{2020A&A...640A..36B,2024ApJS..275....2S} is expected to diminish upon a steep drop of the latter, as this can happen during the RSG phase via pulsation-driven superwinds  \citep{2010ApJ...717L..62Y}.

\section{Conclusions}
\label{sec:summ}

In the present work, we undertook a variability study of 41 post-main-sequence BSGs in the Galaxy with parameters (\logt{}{}, log\,$g$) determined from FEROS spectroscopy  \citep{2010MNRAS.404.1306F} and our own SED modeling to retrieve stellar luminosities. Alongside, the Eddington factor \gammae{}{} of the stars was explored. Our analysis was performed using data from the TESS survey across Sectors $5-66$. We described the photometric time domain by means of three statistical measures; the standard deviation, the coherency parameter $\psi^2$, and the skewness, to assess the degree of scatter, stochasticity and asymmetry of the light curves, respectively. In the frequency domain, we extracted prominent frequencies via iterative pre-whitening and modeled the background signal, indicative of SLF variability, that manifests ubiquitously in the residual spectra.

We report a positive correlation between \logl{}{} and the amplitude of the TESS light curves. Stars with \logl{\lesssim}{5} display independent frequencies that comply with the rotational ones, suggesting rotational modulation as a plausible mechanism for their variability. The majority of these stars display light curves with negative skewness, which has been shown to be associated with line-driven wind instability. For \logl{\gtrsim}{5}, our sample is further populated by the reported \acyg~variables (and presumably more unidentified counterparts of the class), which possess multiperiodic variability with diverse, in some cases time-variant, properties. The irregular light curves of these stars mirror the documented and diverse mechanisms acting, such as mixed-mode oscillations including radial strange modes, winds, and (at least for HD\,77581) binarity.

A significant positive correlation is found between \logt{}{} and the amplitude of the SLF variability, indicating that the ambiguous signal is more pronounced at earlier phases and drops as the stars evolve beyond the main sequence. Essentially, this result contradicts the scenario that SLF variability is excited by sub-surface convection where the opposite effect would be anticipated. Consistent with previous findings, a negative (positive) association is seen between the characteristic timescale of the signal and \logt{}{} (respectively, \gammae{}{}).

We report that the amplitude of SLF variability scales weakly with the absolute $G-$band magnitude for stars fainter than $M_G\sim-6.4$ mag, the limit that here confines the \acyg~variables. Although being nonsignificant, this trend is similar to that for BSGs in the LMC \citep{2019NatAs...3..760B}, which may credit the scenario that the background signal is driven by core-excited IGWs. On the other hand, the discussed trend is vanished within the group of \acyg~variables, with the latter stars being predisposed to lower SLF variability amplitudes. Given that the stellar age appears in this work to be the most potent parameter in regulating the excitation of the signal, we suggest that the suppressed SLF variability of the \acyg~variables could mirror their rather advanced phase as post-RSG stars.


A study over a larger statistical sample would essentially enable a more improved mapping of variability in BSGs onto the evolutionary diagram. Moreover, theoretical insights to the internal or surface processes of \acyg~stars are needed for interpreting their distinctive red noise. It is yet to be determined whether these stars undergo changes in the deep interior that can be assistive to their mass loss at the RSG phase, providing at the same time an explanation of their poorly explained surface abundances.

\begin{acknowledgements}

We thank the anonymous referee for providing a detailed report that greatly improved the quality of this work. The project has received funding from the European Union’s Framework Programme for Research and Innovation Horizon 2020 (2014–2020) under the Marie Skłodowska-Curie grant agreement no. 823734. LC acknowledges financial support from CONICET (PIP 1337) and the University of La Plata (Programa de Incentivos 11/G160). The Astronomical Institute Ondřejov is supported by RVO:67985815. MRD acknowledges support from a CONICET fellowship. This research has made use of the SIMBAD data base, operated at CDS, Strasbourg, France.

\end{acknowledgements}

\begin{appendix}

\section{Supplementary tables}

\begin{table*}
\centering
\caption{\label{tab:sample} List of the 41 studied BSGs in the Galaxy. Coordinates and spectral types are taken from Simbad. The last four columns contain a log of the TESS observations, including the crowding metric for the adopted aperture.}
\renewcommand{\arraystretch}{1.2}
\begin{tabular}{llrrl|cccr}
\hline\hline
\multirow{2}{*}{\#} & \multirow{2}{*}{Star} & RA & DEC & \multirow{2}{*}{Sp. type} & \multirow{2}{*}{TIC} & $T$ & \multirow{2}{*}{\texttt{CROWDSAP$^{(+)}$}} & \multirow{2}{*}{TESS sectors} \\
 & & [deg] & [deg] &  & & [mag] & &  \\
\hline
01 & HD\,51309          & 104.034250 & -17.054239 & B3Ib      & 146908355 & 4.5 & 0.999 & 6,7,33 \\
02 & HD\,64760          & 118.325625 & -48.102933 & B0.5Ib    & 269064487 & 4.4 & 0.983 & 7,8,9,34,35,61,62 \\
03 & HD\,74371\tablefootmark{a}      & 130.487083 & -45.410711 & B6Iab/b   & 285823086 & 5.0 & 0.994 & 8,9,35,36,61,62 \\
04 & HD\,75149          & 131.627250 & -45.912506 & B5Iab     & 28656140  & 5.2 & 0.996 & 8,9,35,36,62 \\
05 & HD\,77581\tablefootmark{a,b}  & 135.528583 & -40.554694 & B0.5Ia    & 191450569 & 6.5 & 0.992 & 8,9,62 \\
06 & HD\,79186          & 137.768333 & -44.867903 & B5Ia      & 74533697  & 4.8 & 0.891 & 8,9,35,36,62,63 \\
07 & HD\,80558\tablefootmark{a}      & 139.676458 & -51.560650 & B6Ia      & 364151367 & 5.3 & 0.996 & 9,10,35,36,62,63 \\
08 & HD\,83183          & 143.611042 & -59.229753 & B5II      & 439431474 & 4.1 & 0.997 & 10,36,62,63,64 \\
09 & HD\,91619\tablefootmark{a}      & 158.355792 & -58.190147 & B6Iab     & 457803940 & 5.6 & 0.994 & 10,36,37,63,64 \\
10 & HD\,92964          & 160.669042 & -59.215758 & B2.5Ia    & 458704469 & 5.2 & 0.987 & 10,36,37,63,64 \\
11 & HD\,93840          & 162.286208 & -46.778286 & B1/2Iab/b & 147215618 & 7.9 & 0.981 & 10,36,37,63 \\
12 & HD\,94493          & 163.312917 & -60.814778 & B0.5Iab/b & 465384125 & 7.3 & 0.938 & 10,11,63,64 \\
13 & HD\,96248          & 166.239875 & -59.858903 & B1Iab     & 305919188 & 6.3 & 0.986 & 10,11,63,64 \\
14 & HD\,99857          & 172.113292 & -66.489275 & B1Ib      & 296059122 & 7.4 & 0.986 & 10,11,37,38,64 \\
15 & HD\,99953\tablefootmark{a}      & 172.313167 & -63.553936 & B2Ia      & 317046099 & 4.8 & 0.998 & 10,11,37,38,64 \\
16 & HD\,103779         & 179.239792 & -63.249089 & B0.5II    & 306562451 & 7.4 & 0.963 & 10,11,37,38 \\
17 & HD\,105071         & 181.473417 & -65.546875 & B8Ia      & 379870355 & 6.1 & 0.990 & 11,37,38,64,65 \\
18 & HD\,106343\tablefootmark{a}     & 183.570542 & -64.408519 & B1.5Ia    & 382515892 & 6.3 & 0.980 & 11,37,38,64,65 \\
19 & HD\,108002         & 186.233167 & -65.211014 & B2Ia/ab   & 450383399 & 6.8 & 0.980 & 11,37,38 \\
20 & HD\,109867         & 189.718167 & -67.193050 & B0.5/1Iab & 327774344 & 6.2 & 0.957 & 11,38,64,65 \\
21 & HD\,111558         & 192.799833 & -69.645208 & B7Ib      & 333977706 & 7.1 & 0.981 & 11,12,38,64 \\
22 & HD\,111973\tablefootmark{b}     & 193.453792 & -60.376239 & B5I       & 436267716 & 5.8 & 0.764 & 11,37,38,64,65 \\
23 & HD\,115842         & 200.201417 & -55.800689 & B0.5Ia    & 256813340 & 5.8 & 0.992 & 11,38,64 \\
24 & HD\,116084         & 200.567875 & -52.182953 & B2Ib      & 401431637 & 5.7 & 0.998 & 11,38,64 \\
25 & HD\,117024         & 202.272583 & -63.874031 & B2Ib      & 314141881 & 7.2 & 0.942 & 11 \\
26 & HD\,119646         & 206.575000 & -62.451600 & B1.5Ib    & 323082858 & 5.3 & 0.996 & 11,38,65 \\
27 & HD\,122879         & 211.604833 & -59.715903 & B0Ia      & 210096523 & 5.3 & 0.990 & 11,38 \\
28 & HD\,125288         & 215.081417 & -56.386494 & B5II      & 413632344 & 4.1 & 0.992 & 11,38,65 \\
29 & HD\,125545         & 215.512583 & -58.291214 & B1Iab/b   & 413965764 & 6.2 & 0.998 & 11,38 \\
30 & HD\,141318         & 237.778333 & -55.055536 & B2III     & 80724706  & 5.9 & 0.842 & 12,39,65 \\
31 & HD\,142758         & 239.848917 & -58.726222 & B1Ia      & 422192335 & 6.9 & 0.925 & 12,39 \\
32 & HD\,148379\tablefootmark{a}     & 247.426375 & -46.243231 & B2Iab     & 226037847 & 4.8 & 0.994 & 12,39,66 \\
33 & HD\,148688\tablefootmark{a}     & 247.924042 & -41.817147 & B1Iaeqp   & 29563900  & 4.9 & 0.955 & 12,39,66 \\
34 & HD\,149038         & 248.520917 & -44.045314 & O9.7Iab   & 228529130 & 4.9 & 0.987 & 39,66 \\
35 & HD\,150168         & 250.417625 & -49.651553 & B1Iab/Ib  & 39177520  & 5.7 & 0.983 & 12,39,66 \\
36 & HD\,150898         & 251.831917 & -58.341442 & B0Ib      & 344237828 & 5.7 & 0.995 & 12,39,66 \\
37 & HD\,152234\tablefootmark{b}     & 253.507667 & -41.806392 & B0.5Ia    & 339565401 & 5.2 & 0.481 & 12,39,66 \\
38 & HD\,152235         & 253.495208 & -41.994328 & B0.5Ia    & 246957312 & 4.3 & 0.997 & 39 \\
39 & HD\,154090         & 256.205625 & -34.122928 & B2Iab     & 41026067  & 4.6 & 0.991 & 12,39,66 \\
40 & HD\,155985         & 259.273083 & -44.778594 & B0.5Ib    & 216802672 & 6.2 & 0.993 & 12,39 \\
41 & HD\,157038\tablefootmark{a}     & 260.663417 & -37.804639 & B3Iap     & 198257590 & 4.4 & 0.991 & 39,66 \\
\hline
\end{tabular}
 \tablefoot{
 \tablefoottext{a}{Star identified as an $\alpha$ Cygni variable \citep{2009A&A...507.1141L,2017ARep...61...80S}.}
  \tablefoottext{b}{Reported in Simbad as a binary/multiple system.}
 }

\end{table*}

\clearpage

\begin{table*}
\small
\caption{\label{tab:cprop} Statistical measures of the TESS light curves and parameters of the SLF variability. Uncertainties are given in parentheses. The middle column contains a flag that indicates when a star is a candidate rotating (cROT) variable.}
\renewcommand{\arraystretch}{1.3}
\begin{tabular}{lccc|c|cccc}
\hline\hline
Star & $\sigma$ [mag] & $\psi^2$ & skw  & cROT & log\,($\alpha_{w}$ [mag]) & log\,($\alpha_{0}$ [mag]) & $\tau$ [d] & $\gamma$ \\
\hline
HD\,51309  & 0.007 (0.001) & 1.67 (0.15) &  0.03 (0.66)  & Y & -5.147 (0.011) & -3.860 (0.004) & 0.055 (0.001) & 4.125 (0.125) \\
HD\,64760  & 0.006 (0.001) & 1.54 (0.06) & -0.13 (0.06)  & Y & -5.642 (0.017) & -3.153 (0.001) & 0.076 (<0.001) & 2.124 (0.011) \\
HD\,74371  & 0.009 (0.002) & 1.84 (0.07) &  0.22 (0.34)  & N & -5.514 (0.003) & -3.720 (0.001) & 0.126 (0.001) & 1.867 (0.009) \\
HD\,75149  & 0.011 (0.003) & 1.91 (0.05) &  0.17 (0.26)  & N & -5.169 (0.003) & -3.654 (0.002) & 0.092 (<0.001) & 5.256 (0.079) \\
HD\,77581  & 0.031 (0.002) & 1.72 (0.16) & -0.07 (0.10)  & Y & -5.192 (0.010) & -3.439 (0.002) & 0.065 (<0.001) & 5.927 (0.126) \\
HD\,79186  & 0.010 (0.002) & 2.04 (0.06) &  0.34 (0.18)  & N & -5.299 (0.004) & -3.106 (0.001) & 0.252 (0.001) & 1.579 (0.007) \\
HD\,80558  & 0.010 (0.002) & 2.19 (0.43) &  0.26 (0.25)  & N & -4.882 (0.031) & -3.493 (0.009) & 0.222 (0.009) & 1.237 (0.036) \\
HD\,83183  & 0.002 (0.001) & 1.53 (0.33) &  0.11 (0.18)  & Y & -5.790 (0.011) & -3.673 (0.002) & 0.073 (<0.001) & 2.990 (0.029) \\
HD\,91619  & 0.013 (0.003) & 1.93 (0.26) &  0.11 (0.12)  & N & -5.207 (0.016) & -4.311 (0.010) & 0.086 (0.003) & 2.596 (0.139) \\
HD\,92964  & 0.014 (0.002) & 1.71 (0.16) & -0.08 (0.35)  & N & -4.607 (0.021) & -3.690 (0.012) & 0.091 (0.004) & 1.927 (0.102) \\
HD\,93840  & 0.011 (0.002) & 1.77 (0.15) &  0.03 (0.05)  & Y & -4.352 (0.024) & -2.937 (0.005) & 0.114 (0.002) & 2.449 (0.068) \\
HD\,94493  & 0.006 (0.001) & 1.59 (0.15) & -0.41 (0.20)  & Y & -4.685 (0.033) & -3.237 (0.006) & 0.092 (0.002) & 2.308 (0.069) \\
HD\,96248  & 0.015 (0.002) & 3.02 (1.11) &  0.24 (0.27)  & Y & -4.308 (0.014) & -3.088 (0.005) & 0.127 (0.002) & 3.962 (0.144) \\
HD\,99857  & 0.005 (0.000) & 1.63 (0.07) & -0.17 (0.06)  & Y & -5.115 (0.012) & -3.110 (0.002) & 0.082 (0.001) & 2.178 (0.022) \\
HD\,99953  & 0.014 (0.002) & 2.16 (0.74) &  0.23 (0.10)  & Y & -5.255 (0.005) & -3.751 (0.003) & 0.074 (<0.001) & 3.798 (0.068) \\
HD\,103779 & 0.007 (<0.001) & 1.62 (0.28) & -0.11 (0.08) & Y & -4.491 (0.022) & -3.229 (0.005) & 0.070 (0.001) & 2.925 (0.091) \\
HD\,105071 & 0.007 (<0.001) & 1.99 (0.47) &  0.10 (0.22) & N & -4.751 (0.016) & -3.585 (0.006) & 0.127 (0.002) & 6.222 (0.383) \\
HD\,106343 & 0.014 (0.002) & 2.28 (0.46) &  0.09 (0.37)  & Y & -4.312 (0.017) & -3.161 (0.006) & 0.103 (0.002) & 4.418 (0.216) \\
HD\,108002 & 0.011 (0.002) & 1.83 (<0.01) & -0.05 (0.24) & Y & -4.206 (0.023) & -2.899 (0.007) & 0.125 (0.002) & 3.631 (0.175) \\
HD\,109867 & 0.012 (0.001) & 2.24 (0.42) & -0.22 (0.50)  & Y & -4.472 (0.015) & -3.533 (0.007) & 0.062 (0.001) & 5.542 (0.334) \\
HD\,111558 & 0.008 (0.001) & 1.94 (0.35) & -0.40 (0.31)  & N & -5.007 (0.009) & -3.718 (0.005) & 0.104 (0.001) & 3.131 (0.090) \\
HD\,111973 & 0.010 (0.003) & 1.91 (0.10) &  0.05 (0.05)  & N & -5.229 (0.005) & -3.662 (0.002) & 0.041 (<0.001) & 2.541 (0.026) \\
HD\,115842 & 0.016 (0.003) & 1.75 (0.03) & -0.31 (0.40)  & Y & -5.506 (0.008) & -3.579 (0.002) & 0.054 (<0.001) & 3.576 (0.040) \\
HD\,116084 & 0.011 (0.001) & 1.80 (0.11) & -0.45 (0.17)  & Y & -5.500 (0.007) & -3.629 (0.002) & 0.072 (<0.001) & 4.095 (0.052) \\
HD\,117024\tablefootmark{a} & 0.003 	       & 1.58 	     & -0.33         & Y & -4.630 (0.025) & -3.458 (0.007) & 0.066 (0.001) & 3.747 (0.184) \\
HD\,119646 & 0.009 (0.001) & 1.78 (0.31) & -0.21 (0.16)  & Y & -4.384 (0.025) & -3.070 (0.008) & 0.121 (0.003) & 2.250 (0.082) \\
HD\,122879 & 0.009 (0.003) & 2.08 (0.31) & -0.18 (0.22)  & N & -5.343 (0.013) & -3.102 (0.001) & 0.082 (<0.001) & 4.269 (0.049) \\
HD\,125288 & 0.002 (0.001) & 1.44 (0.22) & -0.69 (0.57)  & Y & -5.244 (0.007) & -4.192 (0.004) & 0.051 (<0.001) & 5.629 (0.202) \\
HD\,125545 & 0.011 (<0.001) & 1.60 (0.32) &  0.02 (0.26) & Y & -4.476 (0.030) & -3.192 (0.007) & 0.073 (0.001) & 3.206 (0.142) \\
HD\,141318 & 0.001 (<0.001) & 1.34 (0.07) & -0.46 (0.10) & Y & -5.584 (0.004) & -4.013 (0.002) & 0.050 (<0.001) & 1.827 (0.015) \\
HD\,142758 & 0.013 (0.003) & 2.40 (0.95) & -0.32 (0.28)  & N & -4.529 (0.016) & -3.478 (0.006) & 0.088 (0.001) & 5.411 (0.317) \\
HD\,148379 & 0.016 (0.006) & 1.89 (0.03) &  0.10 (0.35)  & N & -5.400 (0.003) & -3.751 (0.002) & 0.118 (0.001) & 1.596 (0.011) \\
HD\,148688 & 0.012 (0.001) & 1.98 (0.09) & -0.09 (0.28)  & N & -5.096 (0.002) & -3.609 (0.002) & 0.106 (0.001) & 2.604 (0.026) \\
HD\,149038 & 0.010 (0.002) & 1.61 (<0.01) &  0.17 (0.16) & Y & -5.238 (0.011) & -3.098 (0.001) & 0.064 (<0.001) & 3.022 (0.026) \\
HD\,150168 & 0.007 (0.001) & 1.73 (0.05) & -0.44 (0.16)  & Y & -5.319 (0.015) & -3.093 (0.002) & 0.074 (0.001) & 2.029 (0.017) \\
HD\,150898 & 0.006 (0.001) & 1.58 (0.08) &  0.00 (0.18)  & Y & -5.339 (0.012) & -3.112 (0.002) & 0.066 (<0.001) & 2.297 (0.016) \\
HD\,152234 & 0.009 (0.002) & 1.70 (0.20) & -0.24 (0.44)  & Y & -5.018 (0.020) & -3.347 (0.004) & 0.077 (0.001) & 3.214 (0.076) \\
HD\,152235\tablefootmark{a} & 0.014 	       & 1.92        &  0.21         & N & -5.312 (0.004) & -3.698 (0.002) & 0.063 (<0.001) & 3.280 (0.034) \\
HD\,154090 & 0.014 (0.004) & 1.74 (0.16) & -0.07 (0.09)  & Y & -5.245 (0.006) & -3.478 (0.002) & 0.065 (<0.001) & 4.483 (0.058) \\
HD\,155985 & 0.006 (0.001) & 1.54 (0.13) &  0.08 (0.32)  & Y & -4.399 (0.024) & -3.294 (0.007) & 0.060 (0.001) & 3.468 (0.166) \\
HD\,157038 & 0.010 (0.001) & 1.79 (0.14) & -0.13 (0.14)  & N & -5.417 (0.002) & -3.922 (0.002) & 0.083 (<0.001) & 2.621 (0.023) \\
\hline
\end{tabular}
\tablefoot{
\tablefoottext{a}{Data studied from a single TESS sector.}
}

\end{table*}

\clearpage

\begin{table*}
\centering
\caption{\label{tab:sed_prop} Eddington factors \gammae{}{} and SED-based scaling factors $C$ of the sample BSGs. The last two columns contain luminosities and the absolute $G-$band magnitudes, which were calculated using the distance values from $Gaia$ (third column).}
\renewcommand{\arraystretch}{1.4}
\begin{tabular}{lc|c|c|cc}
\hline\hline
Star & \gammae{}{} & $D_{Gaia}$ [pc] & log\,$C$ & \logl{}{} & M$_{G}$ [mag]\\
\hline
HD\,51309  & 0.18 & 1108$_{-214}^{+408}$ & -1.49 & 4.84$_{-0.19}^{+0.27}$ & -6.06$_{+0.47}^{-0.68}$ \\
HD\,64760  & 0.19 &  659$_{ -53}^{ +58}$ & -1.67 & 4.91$_{-0.07}^{+0.07}$ & -5.17$_{+0.18}^{-0.19}$ \\
HD\,74371  & 0.31 & 1718$_{-141}^{+177}$ & -1.43 & 5.07$_{-0.07}^{+0.09}$ & -6.87$_{+0.19}^{-0.21}$ \\
HD\,75149  & 0.31 & 1466$_{-123}^{+121}$ & -1.47 & 5.16$_{-0.09}^{+0.06}$ & -6.58$_{+0.19}^{-0.17}$ \\
HD\,77581  & 0.47 & 1969$_{- 68}^{+ 61}$ & -1.79 & 5.66$_{-0.04}^{+0.02}$ & -6.51$_{+0.08}^{-0.07}$ \\
HD\,79186  & 0.39 & 1677$_{-201}^{+212}$ & -1.41 & 5.29$_{-0.10}^{+0.11}$ & -7.14$_{+0.28}^{-0.26}$ \\
HD\,80558  & 0.43 & 1862$_{ -90}^{ +80}$ & -1.34 & 5.27$_{-0.04}^{+0.03}$ & -7.27$_{+0.11}^{-0.09}$ \\
HD\,83183  & 0.12 &  593$_{ -56}^{ +55}$ & -1.38 & 4.44$_{-0.09}^{+0.07}$ & -5.16$_{+0.21}^{-0.19}$ \\
HD\,91619  & 0.44 & 2319$_{-100}^{+144}$ & -1.48 & 5.18$_{-0.03}^{+0.06}$ & -7.10$_{+0.10}^{-0.13}$ \\
HD\,92964  & 0.45 & 2130$_{-238}^{+288}$ & -1.44 & 5.50$_{-0.10}^{+0.11}$ & -7.46$_{+0.26}^{-0.28}$ \\
HD\,93840  & 0.26 & 2649$_{-231}^{+241}$ & -2.22 & 4.64$_{-0.08}^{+0.08}$ & -4.82$_{+0.20}^{-0.19}$ \\
HD\,94493  & 0.26 & 2167$_{ -94}^{+109}$ & -2.13 & 4.85$_{-0.04}^{+0.04}$ & -5.11$_{+0.10}^{-0.11}$ \\
HD\,96248  & 0.44 & 2426$_{-113}^{+122}$ & -1.81 & 5.27$_{-0.05}^{+0.04}$ & -6.43$_{+0.10}^{-0.11}$ \\
HD\,99857  & 0.18 & 1794$_{ -57}^{ +75}$ & -2.08 & 4.63$_{-0.03}^{+0.04}$ & -4.75$_{+0.07}^{-0.09}$ \\
HD\,99953  & 0.43 & 2490$_{-127}^{+165}$ & -1.64 & 5.37$_{-0.05}^{+0.05}$ & -6.90$_{+0.11}^{-0.14}$ \\
HD\,103779 & 0.30 & 2039$_{ -96}^{+134}$ & -2.13 & 4.90$_{-0.04}^{+0.06}$ & -5.05$_{+0.10}^{-0.14}$ \\
HD\,105071 & 0.24 & 2187$_{ -94}^{ +96}$ & -1.62 & 4.73$_{-0.03}^{+0.05}$ & -6.24$_{+0.10}^{-0.09}$ \\
HD\,106343 & 0.39 & 2379$_{-125}^{+144}$ & -1.81 & 5.30$_{-0.05}^{+0.05}$ & -6.51$_{+0.12}^{-0.13}$ \\
HD\,108002 & 0.32 & 2391$_{ -93}^{+103}$ & -1.93 & 5.07$_{-0.03}^{+0.04}$ & -5.88$_{+0.09}^{-0.09}$ \\
HD\,109867 & 0.33 & 2552$_{-172}^{+219}$ & -1.86 & 5.43$_{-0.05}^{+0.08}$ & -6.56$_{+0.15}^{-0.18}$ \\
HD\,111558 & 0.16 & 4594$_{-516}^{+589}$ & -1.86 & 4.90$_{-0.12}^{+0.09}$ & -6.69$_{+0.26}^{-0.26}$ \\
HD\,111973 & 0.25 & 1948$_{-151}^{+203}$ &  -1.6 & 5.15$_{-0.07}^{+0.08}$ & -6.57$_{+0.17}^{-0.21}$ \\
HD\,115842 & 0.51 & 1682$_{ -91}^{ +84}$ & -1.71 & 5.57$_{-0.06}^{+0.03}$ & -6.53$_{+0.12}^{-0.11}$ \\
HD\,116084 & 0.29 & 1958$_{-147}^{+161}$ & -1.67 & 5.04$_{-0.07}^{+0.06}$ & -6.36$_{+0.17}^{-0.17}$ \\
HD\,117024 & 0.16 & 1984$_{ -92}^{ +84}$ & -1.98 & 4.47$_{-0.05}^{+0.02}$ & -4.96$_{+0.10}^{-0.09}$ \\
HD\,119646 & 0.23 & 1721$_{ -73}^{ +84}$ & -1.84 & 4.73$_{-0.04}^{+0.04}$ & -5.40$_{+0.09}^{-0.10}$ \\
HD\,122879 & 0.52 & 2274$_{-195}^{+179}$ & -1.91 & 5.59$_{-0.08}^{+0.06}$ & -6.44$_{+0.20}^{-0.16}$ \\
HD\,125288 & 0.08 &  438$_{ -30}^{ +51}$ & -1.34 & 4.13$_{-0.06}^{+0.09}$ & -4.50$_{+0.16}^{-0.24}$ \\
HD\,125545 & 0.25 & 2171$_{ -89}^{ +84}$ & -2.01 & 4.87$_{-0.03}^{+0.04}$ & -5.29$_{+0.09}^{-0.08}$ \\
HD\,141318 & 0.11 &  596$_{ -19}^{ +14}$ & -1.73 & 4.09$_{-0.02}^{+0.02}$ & -3.71$_{+0.07}^{-0.05}$ \\
HD\,142758 & 0.38 & 3156$_{-161}^{+172}$ & -1.86 & 5.19$_{-0.05}^{+0.04}$ & -6.49$_{+0.11}^{-0.12}$ \\
HD\,148379 & 0.63 & 1585$_{-168}^{+204}$ & -1.29 & 5.70$_{-0.10}^{+0.10}$ & -7.66$_{+0.24}^{-0.26}$ \\
HD\,148688 & 0.49 & 1348$_{-154}^{+215}$ & -1.44 & 5.60$_{-0.11}^{+0.12}$ & -6.97$_{+0.26}^{-0.32}$ \\
HD\,149038 & 0.40 &  896$_{ -84}^{ +91}$ & -1.67 & 5.33$_{-0.08}^{+0.09}$ & -5.75$_{+0.21}^{-0.21}$ \\
HD\,150168 & 0.20 & 1015$_{ -66}^{ +75}$ &  -1.8 & 4.94$_{-0.05}^{+0.07}$ & -5.54$_{+0.15}^{-0.15}$ \\
HD\,150898 & 0.29 & 1118$_{ -83}^{+112}$ & -1.87 & 4.99$_{-0.06}^{+0.09}$ & -5.22$_{+0.17}^{-0.21}$ \\
HD\,152234 & 0.42 & 1935$_{-575}^{+586}$ & -1.65 & 5.86$_{-0.30}^{+0.24}$ & -7.16$_{+0.77}^{-0.57}$ \\
HD\,152235 & 0.40 & 1618$_{ -94}^{+133}$ &  -1.6 & 5.55$_{-0.06}^{+0.06}$ & -6.73$_{+0.13}^{-0.17}$ \\
HD\,154090 & 0.37 & 1116$_{-137}^{+131}$ & -1.45 & 5.54$_{-0.12}^{+0.09}$ & -6.76$_{+0.28}^{-0.24}$ \\
HD\,155985 & 0.25 & 1077$_{ -35}^{ +38}$ & -1.83 & 4.82$_{-0.03}^{+0.03}$ & -4.96$_{+0.07}^{-0.08}$ \\
HD\,157038 & 0.46 & 2168$_{-102}^{+134}$ & -1.34 & 5.74$_{-0.05}^{+0.04}$ & -7.89$_{+0.10}^{-0.13}$ \\
\hline
\end{tabular}
\end{table*}

\clearpage

\section{Supplementary figures.}
\label{app:supp_fig}

In this section, we provide supplementary figures following the Figs. \ref{fig:lc_0}, \ref{fig:ls_0}, and \ref{fig:seds_0}.    

\begin{figure*}
   \centering
   \includegraphics[width=16cm]{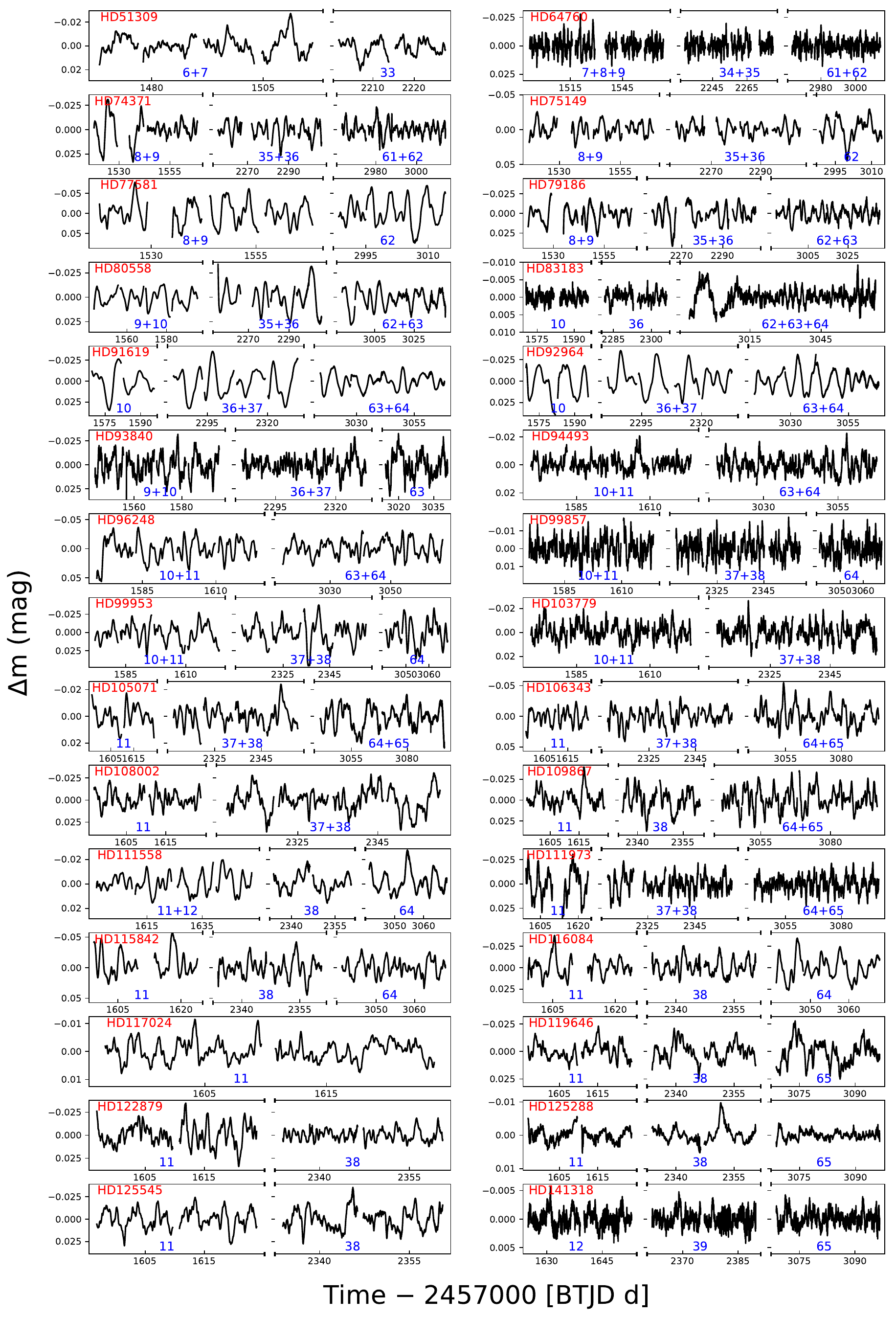}
      \caption{\label{app:lc_1} TESS photometry as in Fig. \ref{fig:lc_0}, for the stars $1-30$ from Table \ref{tab:sample}.        }
\end{figure*}

\begin{figure*}
   \centering
   \includegraphics[width=16.7cm]{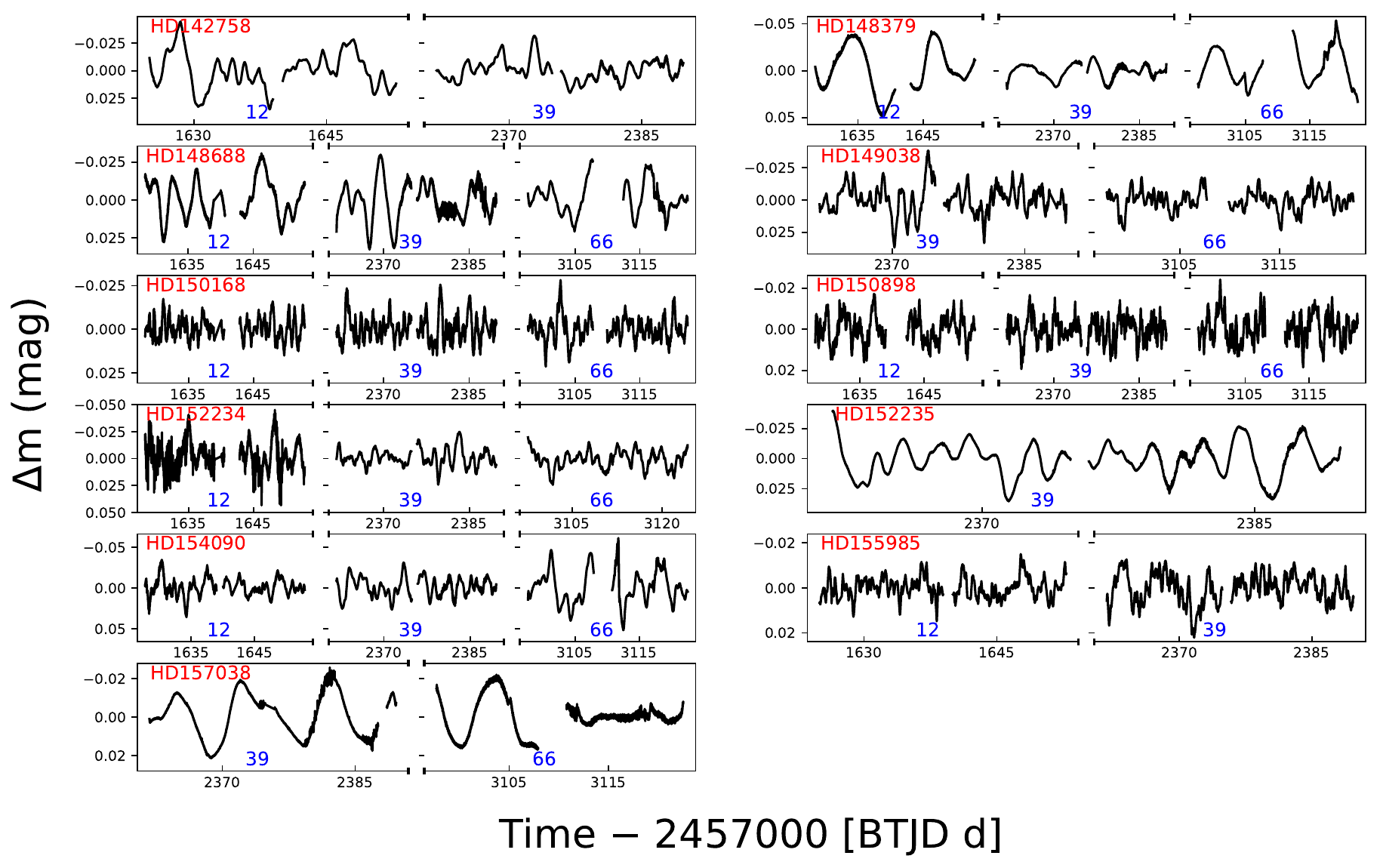}
      \caption{\label{app:lc_2}  TESS photometry as in Fig. \ref{fig:lc_0}, for the stars $31-41$ from Table \ref{tab:sample}.       }
\end{figure*}
   
\begin{figure*}
   \centering
   \includegraphics[width=16.7cm]{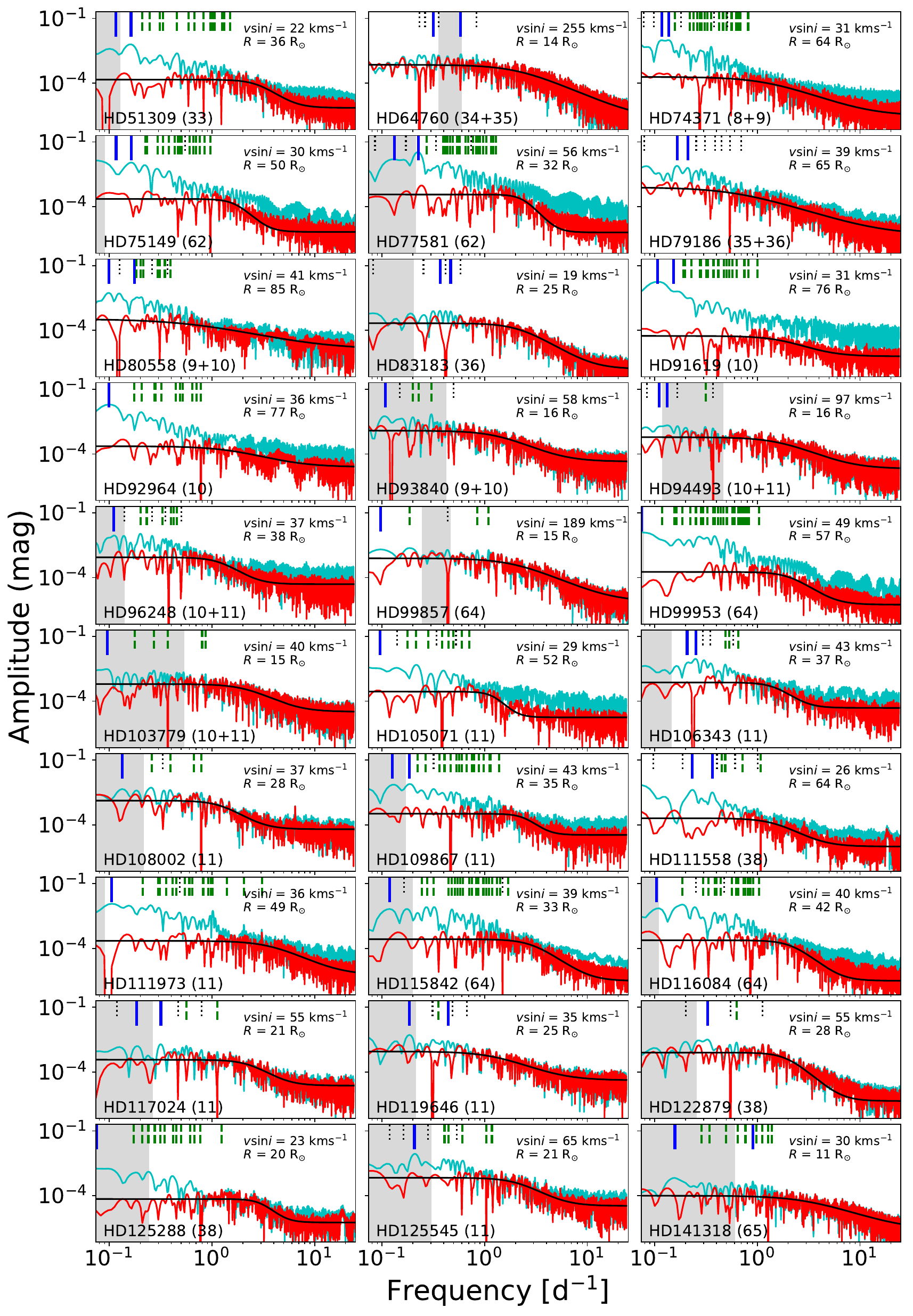}
      \caption{\label{app:ls_1} Frequency spectra as in Fig. \ref{fig:ls_0}, for the stars $1-30$ from Table \ref{tab:sample}. }        
\end{figure*}

\begin{figure*}
   \centering
   \includegraphics[width=16.7cm]{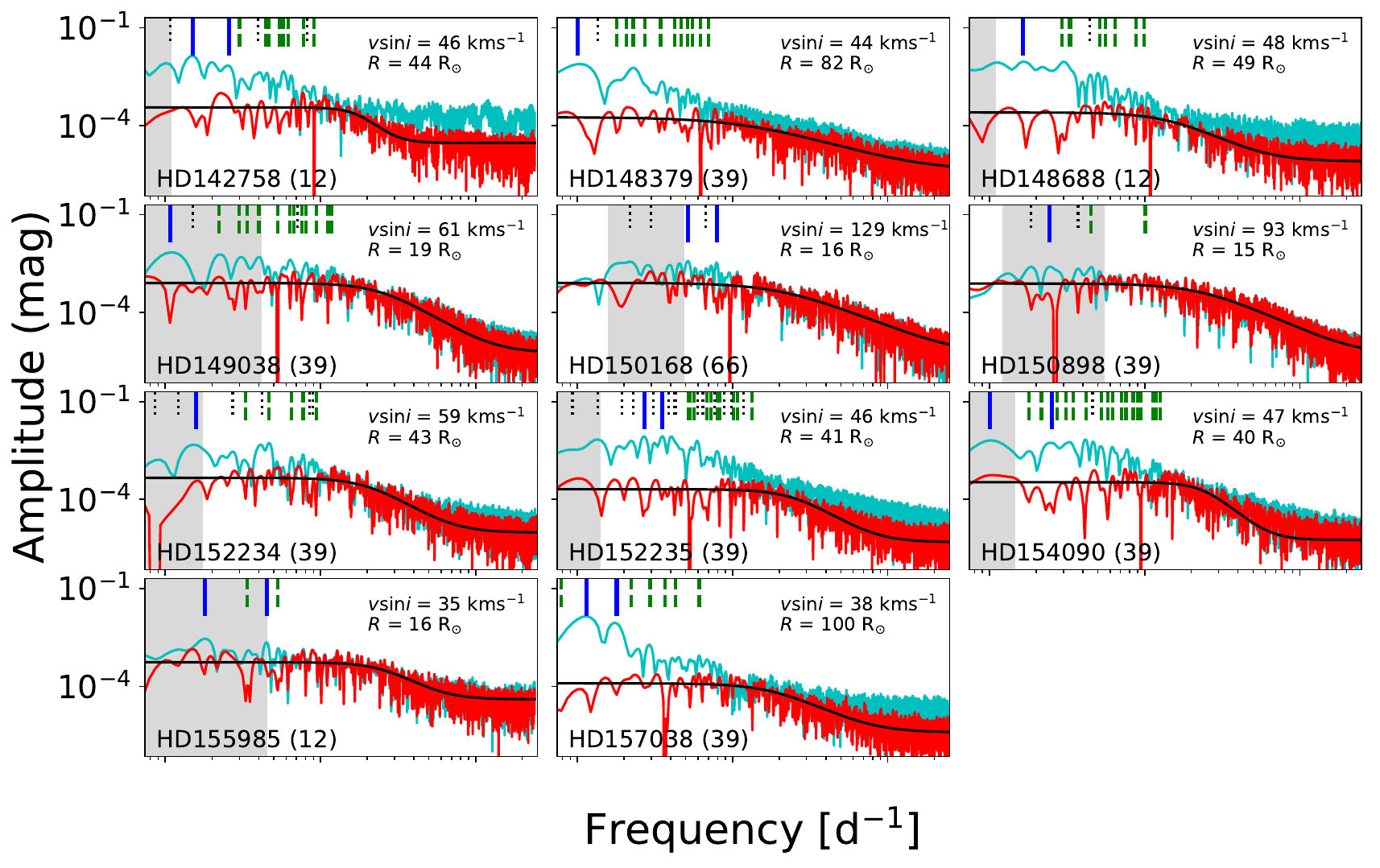}
      \caption{\label{{app:ls_2}} Frequency spectra as in Fig. \ref{fig:ls_0}, for the stars $31-41$ from Table \ref{tab:sample}.        }
\end{figure*}

\label{app:sed}
   \begin{figure*}
   \centering
   \includegraphics[width=17cm]{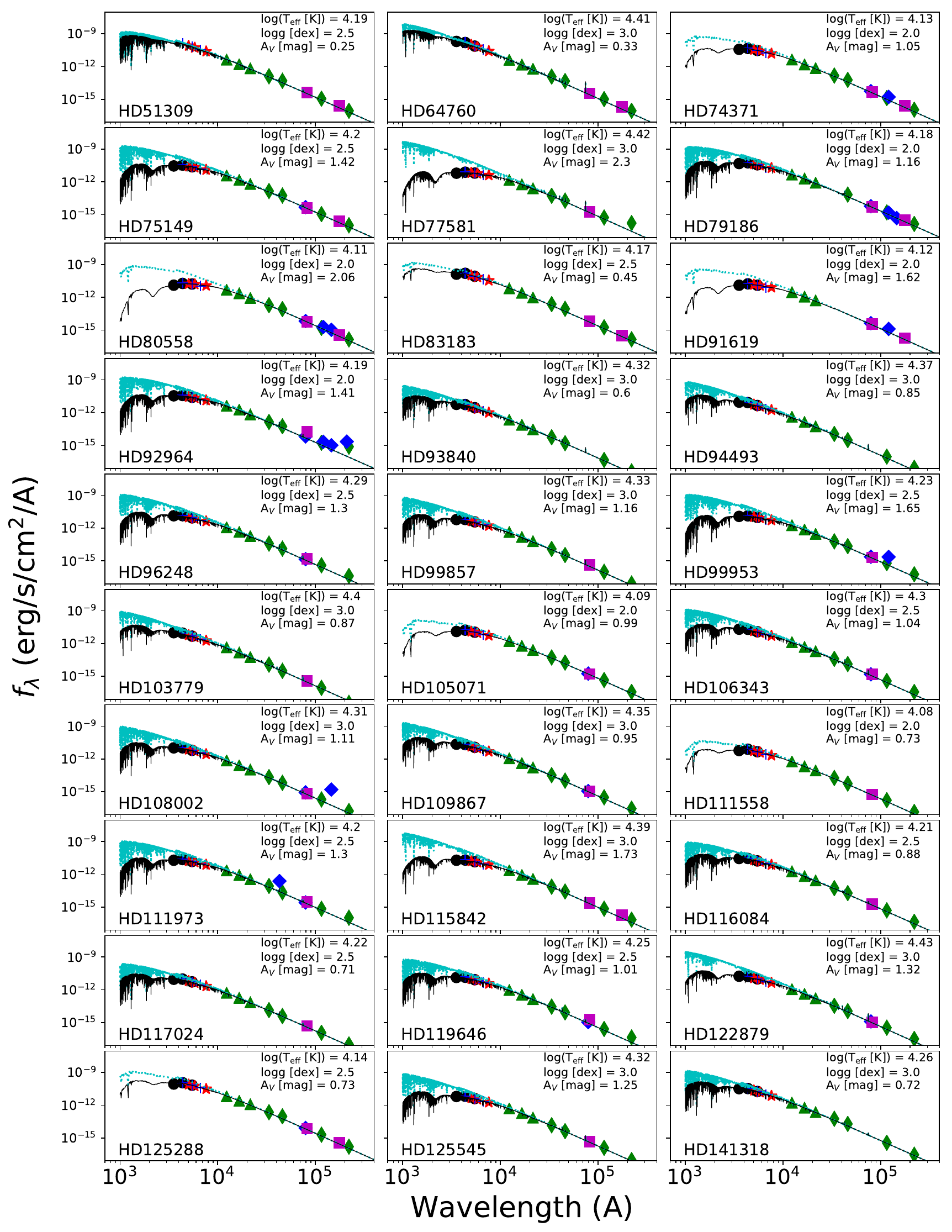}
      \caption{\label{{app:sed_1}} Spectral energy distributions and their fitting models as in Fig. \ref{fig:seds_0}, for the stars $1-30$ from Table \ref{tab:sample}. }      
   \end{figure*}

\begin{figure*}
   \centering
   \includegraphics[width=17cm]{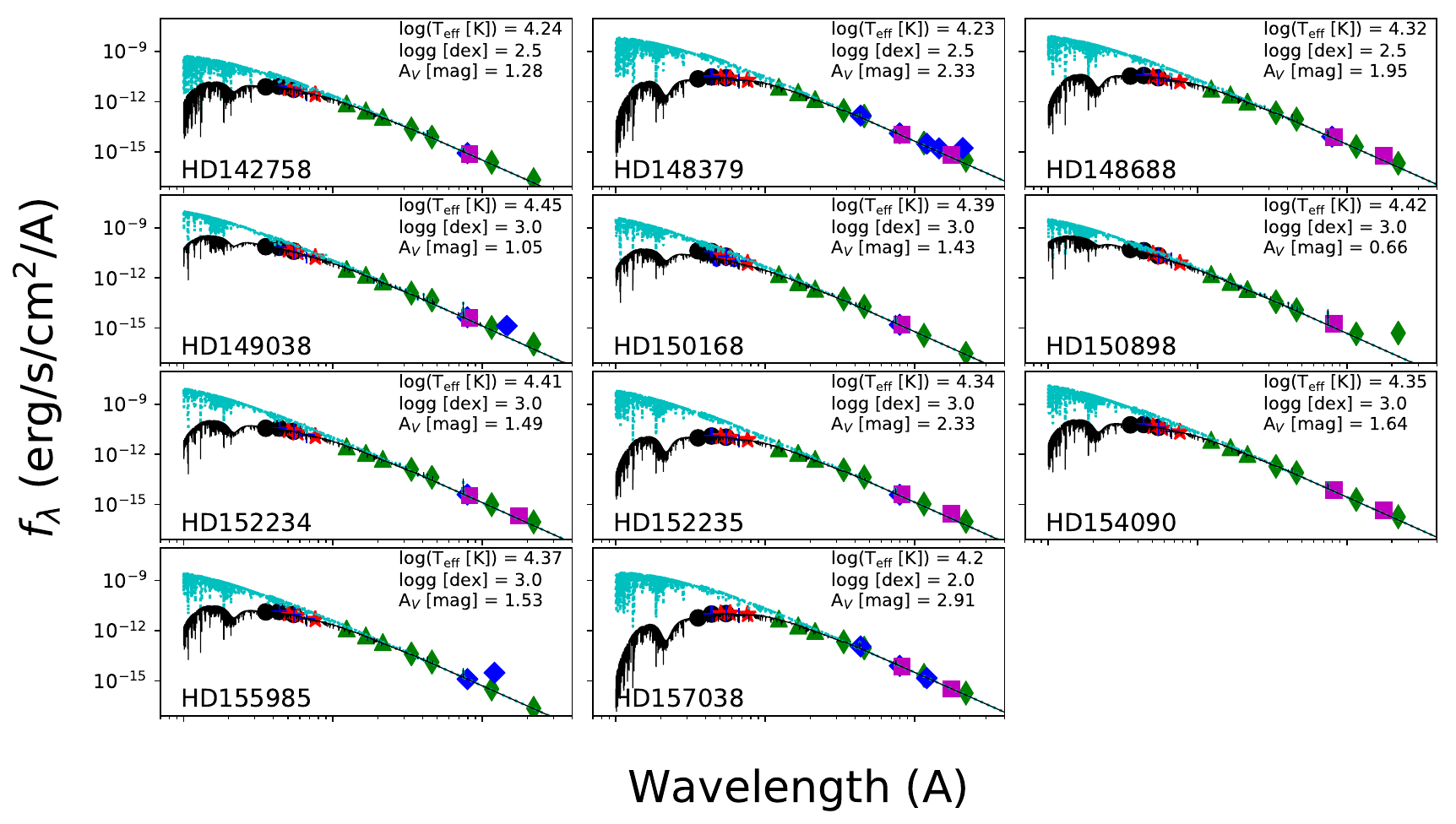}
      \caption{\label{{app:sed_2}} Spectral energy distributions and their fitting models as in Fig. \ref{fig:seds_0}, for the stars $31-41$ from Table \ref{tab:sample}.          }         
   \end{figure*}
   
\clearpage

\section{Independent frequencies}
\label{app:supp_ffreq}

Independent frequencies of the studied BSGs that were extracted via the iterative pre-whitening process. The resolution of the process over the different windows (consisting of single or stitched consecutive sectors) is determined by the Rayleigh criterion. The S/N values are calculated with respect to the adopted model of the SLF variability.

\clearpage

{\centering
\twocolumn
\tablehead{
\hline\hline
\multirow{2}{*}{Star} & \multirow{2}{*}{Sector(s)} & $f_{i}$ & $A_{i}$ & \multirow{2}{*}{S/N}\\
& & [d$^{-1}$] & [mmag] & \\
\hline}
\tabletail{\hline}
\tablecaption{\label{app:ffreq} Independent frequencies of the studied BSGs that were extracted from the different observing windows.}
\begin{supertabular}{lcccc}
HD51309 & 33 & 0.1627 & 6.3 & 43\\ 
 &  & 0.1162 & 2.7 & 18\\ 
HD51309 & 6+7 & 0.0838 & 5.5 & 7\\ 
HD64760 & 34+35 & 0.5846 & 2.2 & 3\\ 
 &  & 0.3178 & 2.1 & 3\\ 
HD64760 & 7+8+9 & 0.4164 & 4.1 & 6\\ 
 &  & 0.3644 & 2.4 & 4\\ 
 &  & 0.4347 & 1.7 & 2\\ 
HD64760 & 61+62 & 0.3868 & 1.9 & 3\\ 
 &  & 0.6236 & 1.8 & 3\\ 
HD74371 & 35+36 & 0.1310 & 5.6 & 15\\ 
HD74371 & 61+62 & 0.2052 & 4.0 & 6\\ 
 &  & 0.2486 & 3.4 & 5\\ 
HD74371 & 8+9 & 0.1377 & 9.8 & 51\\ 
 &  & 0.1181 & 10.3 & 53\\ 
HD75149 & 35+36 & 0.1867 & 6.4 & 17\\ 
 &  & 0.3098 & 4.1 & 12\\ 
 &  & 0.2184 & 4.3 & 12\\ 
HD75149 & 8+9 & 0.1535 & 6.0 & 5\\ 
 &  & 0.2046 & 5.3 & 5\\ 
HD75149 & 62 & 0.1634 & 10.1 & 44\\ 
 &  & 0.1167 & 6.5 & 28\\ 
HD77581 & 62 & 0.2256 & 34.0 & 91\\ 
 &  & 0.1323 & 18.2 & 49\\ 
HD77581 & 8+9 & 0.2203 & 28.9 & 55\\ 
 &  & 0.1535 & 10.5 & 20\\ 
HD79186 & 8+9 & 0.1141 & 6.2 & 11\\ 
 &  & 0.1653 & 6.6 & 12\\ 
HD79186 & 62+63 & 0.1809 & 4.1 & 7\\ 
 &  & 0.3079 & 4.2 & 8\\ 
HD79186 & 35+36 & 0.1668 & 7.8 & 11\\ 
 &  & 0.2105 & 7.9 & 11\\ 
HD80558 & 35+36 & 0.1412 & 11.1 & 20\\ 
 &  & 0.1177 & 6.5 & 11\\ 
HD80558 & 62+63 & 0.0991 & 7.6 & 25\\ 
 &  & 0.1640 & 4.9 & 16\\ 
HD80558 & 9+10 & 0.0994 & 5.5 & 17\\ 
 &  & 0.1759 & 3.5 & 12\\ 
HD83183 & 62+63+64 & 0.0788 & 1.2 & 4\\ 
HD83183 & 36 & 0.3684 & 0.8 & 3\\ 
 &  & 0.4666 & 0.7 & 3\\ 
HD83183 & 10 & 0.4676 & 0.6 & 4\\ 
 &  & 0.3804 & 0.5 & 3\\ 
HD91619 & 63+64 & 0.1267 & 6.0 & 57\\ 
 &  & 0.0969 & 4.7 & 44\\ 
HD91619 & 36+37 & 0.1239 & 14.0 & 108\\ 
 &  & 0.0813 & 8.8 & 67\\ 
HD91619 & 10 & 0.1070 & 17.8 & 322\\ 
 &  & 0.1529 & 7.5 & 135\\ 
HD92964 & 63+64 & 0.1342 & 10.1 & 46\\ 
 &  & 0.1118 & 6.2 & 28\\ 
HD92964 & 10 & 0.0994 & 19.6 & 85\\ 
HD92964 & 36+37 & 0.1355 & 11.8 & 71\\ 
 &  & 0.0813 & 8.9 & 53\\ 
HD93840 & 9+10 & 0.1076 & 5.4 & 4\\ 
HD93840 & 63 & 0.1056 & 12.6 & 7\\ 
 &  & 0.4675 & 4.9 & 3\\ 
HD93840 & 36+37 & 0.1045 & 5.2 & 5\\ 
 &  & 0.2401 & 3.7 & 4\\ 
HD94493 & 63+64 & 0.1789 & 3.1 & 4\\ 
 &  & 0.3243 & 2.8 & 3\\ 
HD94493 & 10+11 & 0.1324 & 2.1 & 3\\ 
 &  & 0.1103 & 2.1 & 3\\ 
HD96248 & 10+11 & 0.1103 & 8.9 & 10\\ 
HD96248 & 63+64 & 0.1401 & 7.6 & 7\\ 
 &  & 0.1022 & 5.5 & 5\\ 
HD99857 & 37+38 & 0.6837 & 2.2 & 4\\ 
 &  & 0.1803 & 1.6 & 3\\ 
HD99857 & 64 & 0.0966 & 2.3 & 2\\ 
HD99857 & 10+11 & 0.3678 & 2.0 & 3\\ 
 &  & 0.1434 & 1.8 & 3\\ 
HD99953 & 64 & 0.0743 & 12.3 & 67\\ 
HD99953 & 37+38 & 0.1127 & 11.6 & 18\\ 
 &  & 0.0751 & 8.4 & 13\\ 
HD99953 & 10+11 & 0.1045 & 6.9 & 11\\ 
 &  & 0.2555 & 6.4 & 10\\ 
HD103779 & 37+38 & 0.1465 & 3.4 & 4\\ 
 &  & 0.4808 & 2.5 & 3\\ 
HD103779 & 10+11 & 0.0956 & 3.1 & 4\\ 
HD105071 & 37+38 & 0.1014 & 4.2 & 23\\ 
HD105071 & 11 & 0.0962 & 3.9 & 14\\ 
HD105071 & 64+65 & 0.1190 & 4.8 & 8\\ 
 &  & 0.3101 & 3.3 & 5\\ 
HD106343 & 37+38 & 0.1013 & 8.8 & 15\\ 
HD106343 & 11 & 0.2517 & 9.2 & 12\\ 
 &  & 0.2072 & 6.4 & 8\\ 
HD106343 & 64+65 & 0.1910 & 9.2 & 4\\ 
 &  & 0.1513 & 7.3 & 3\\ 
HD108002 & 37+38 & 0.1352 & 5.5 & 4\\ 
 &  & 0.3043 & 4.9 & 4\\ 
HD108002 & 11 & 0.1332 & 4.6 & 3\\ 
HD109867 & 64+65 & 0.3387 & 6.4 & 9\\ 
 &  & 0.2090 & 6.2 & 8\\ 
HD109867 & 11 & 0.1850 & 7.8 & 23\\ 
 &  & 0.1258 & 6.2 & 18\\ 
HD109867 & 38 & 0.1098 & 11.7 & 13\\ 
HD111558 & 64 & 0.1338 & 6.7 & 37\\ 
 &  & 0.2230 & 5.7 & 32\\ 
HD111558 & 11+12 & 0.1035 & 3.7 & 9\\ 
 &  & 0.2571 & 3.0 & 7\\ 
HD111558 & 38 & 0.2324 & 4.1 & 20\\ 
 &  & 0.3674 & 2.2 & 10\\ 
HD111973 & 11 & 0.1058 & 11.8 & 52\\ 
HD111973 & 37+38 & 1.0212 & 5.3 & 14\\ 
 &  & 0.2858 & 4.8 & 8\\ 
HD111973 & 64+65 & 1.0196 & 4.9 & 10\\ 
 &  & 0.2774 & 4.4 & 5\\ 
 &  & 0.2126 & 3.3 & 4\\ 
HD115842 & 64 & 0.1189 & 7.2 & 26\\ 
HD115842 & 38 & 0.2473 & 11.5 & 32\\ 
 &  & 0.4271 & 8.1 & 23\\ 
HD115842 & 11 & 0.1628 & 16.2 & 23\\ 
HD116084 & 64 & 0.1040 & 7.7 & 32\\ 
HD116084 & 11 & 0.2017 & 9.4 & 27\\ 
 &  & 0.2985 & 6.3 & 18\\ 
HD116084 & 38 & 0.2398 & 6.3 & 18\\ 
 &  & 0.3072 & 4.6 & 13\\ 
HD117024 & 11 & 0.3183 & 1.9 & 5\\ 
 &  & 0.1850 & 1.6 & 4\\ 
HD119646 & 11 & 0.1850 & 4.8 & 5\\ 
 &  & 0.4441 & 2.8 & 3\\ 
HD119646 & 38 & 0.1350 & 6.4 & 12\\ 
 &  & 0.2099 & 3.6 & 7\\ 
HD119646 & 65 & 0.1078 & 8.8 & 10\\ 
 &  & 0.1724 & 3.7 & 4\\ 
HD122879 & 38 & 0.3297 & 3.2 & 4\\ 
HD122879 & 11 & 0.2739 & 6.9 & 4\\ 
 &  & 0.4811 & 5.9 & 3\\ 
HD125288 & 38 & 0.0750 & 1.8 & 25\\ 
HD125288 & 65 & 0.1083 & 0.6 & 3\\ 
HD125288 & 11 & 0.1258 & 1.2 & 10\\ 
 &  & 0.3183 & 0.8 & 7\\ 
HD125545 & 11 & 0.2072 & 8.7 & 12\\ 
HD125545 & 38 & 0.1125 & 6.0 & 4\\ 
HD141318 & 65 & 0.1578 & 0.7 & 6\\ 
 &  & 0.9037 & 0.5 & 5\\ 
HD141318 & 39 & 0.2863 & 0.5 & 2\\ 
 &  & 0.5297 & 0.4 & 2\\ 
HD141318 & 12 & 0.2506 & 0.9 & 11\\ 
 &  & 0.1790 & 0.6 & 8\\ 
HD142758 & 12 & 0.1508 & 13.8 & 38\\ 
 &  & 0.2585 & 7.9 & 21\\ 
HD142758 & 39 & 0.1648 & 5.0 & 8\\ 
 &  & 0.2795 & 4.2 & 6\\ 
HD148379 & 66 & 0.1131 & 20.0 & 64\\ 
HD148379 & 12 & 0.1637 & 15.2 & 49\\ 
HD148379 & 39 & 0.1003 & 7.7 & 42\\ 
HD148688 & 39 & 0.2435 & 10.3 & 27\\ 
 &  & 0.1790 & 7.7 & 19\\ 
HD148688 & 66 & 0.1212 & 8.8 & 20\\ 
HD148688 & 12 & 0.1637 & 9.1 & 36\\ 
HD149038 & 39 & 0.1074 & 6.9 & 8\\ 
HD149038 & 66 & 0.1778 & 5.7 & 7\\ 
HD150168 & 66 & 0.5172 & 3.8 & 4\\ 
 &  & 0.7919 & 3.8 & 5\\ 
HD150168 & 39 & 0.7878 & 4.7 & 10\\ 
 &  & 0.3724 & 4.2 & 8\\ 
HD150168 & 12 & 0.8294 & 4.0 & 9\\ 
 &  & 0.3659 & 2.7 & 6\\ 
HD150898 & 66 & 0.1286 & 3.9 & 6\\ 
HD150898 & 39 & 0.2434 & 2.6 & 3\\ 
HD150898 & 12 & 0.1850 & 4.1 & 5\\ 
 &  & 0.4264 & 4.0 & 5\\ 
HD152234 & 39 & 0.1577 & 4.6 & 10\\ 
HD152234 & 12 & 0.2277 & 7.0 & 4\\ 
 &  & 0.5285 & 6.3 & 3\\ 
HD152234 & 66 & 0.1055 & 5.0 & 13\\ 
HD152235 & 39 & 0.3509 & 8.7 & 42\\ 
 &  & 0.2722 & 8.0 & 39\\ 
HD154090 & 12 & 0.3740 & 7.2 & 5\\ 
 &  & 0.1671 & 5.9 & 4\\ 
HD154090 & 39 & 0.1010 & 6.2 & 18\\ 
 &  & 0.2524 & 5.2 & 15\\ 
HD154090 & 66 & 0.1127 & 16.7 & 32\\ 
 &  & 0.2897 & 13.9 & 26\\ 
HD155985 & 12 & 0.1795 & 2.9 & 5\\ 
 &  & 0.4523 & 2.0 & 3\\ 
HD155985 & 39 & 0.1935 & 3.7 & 6\\ 
 &  & 0.3225 & 3.4 & 5\\ 
HD157038 & 39 & 0.1149 & 14.0 & 113\\ 
 &  & 0.1795 & 6.2 & 50\\ 
HD157038 & 66 & 0.1449 & 9.3 & 44\\ 
\hline
\end{supertabular}}

\end{appendix}

\end{document}